\documentclass{article}

\usepackage{arxiv}

\usepackage[utf8]{inputenc} 
\usepackage[T1]{fontenc}    
\usepackage{hyperref}       
\usepackage{url}            
\usepackage{booktabs}       
\usepackage{amsfonts}       
\usepackage{nicefrac}       
\usepackage{microtype}      
\usepackage{lipsum}
\usepackage{graphicx}
\usepackage{fancyhdr}
\usepackage[most]{tcolorbox}
\usepackage{pgfplots}
\pgfplotsset{compat=1.18}
\usepackage{longtable}
\graphicspath{ {./images/} }

\usepackage{enumitem}
\usepackage{booktabs}
\usepackage{tabularx}
\usepackage{colortbl}
\usepackage{xcolor}
\usepackage{array}
\usepackage{multirow}
\usepackage{caption}
\usepackage{titlesec}
\usepackage[table]{xcolor}

\definecolor{groupbg}{gray}{0.85}
 
\newcolumntype{S}{>{\arraybackslash}p{1.7cm}}
\newcolumntype{M}{>{\arraybackslash}p{1.9cm}}

\newcounter{pattern}
\newcommand{\pattern}[1]{%
  \refstepcounter{pattern}%
  \noindent{\large\textsc{Pattern \thepattern{} - \textsc{#1}}}%
  \addvspace{5pt}
}

\usepackage{sectsty}
\subsectionfont{\large}

\title{A Catalog of User Authentication Patterns}

\author{
  Alex R. Mattukat \\
  Research Group Software Construction\\
  RWTH Aachen University\\
  Ahornstraße 55 \\
  Aachen, Germany \\
  \texttt{mattukat@swc.rwth-aachen.de}
  \And
  Horst Lichter \\
  Research Group Software Construction\\
  RWTH Aachen University\\
  Ahornstraße 55 \\
  Aachen, Germany \\
  \texttt{lichter@swc.rwth-aachen.de} \\
}

\begin{document}
\tiny
This is the \textbf{peer-reviewed, accepted version} of a paper. It will appear in the proceedings of the 33rd Conference on Pattern Languages of Programs, People, and Practices (\textbf{PLoP 2026}). The final published version will be available from the Association for Computing Machinery (ACM) via the ACM Digital Library.
© 2026 ACM. Personal use of this material is permitted. Permission from ACM must be obtained for all other uses, in any current or future media, including reprinting or republishing this material for advertising or promotional purposes, creating new collective works, resale or redistribution to servers or lists, or reuse of any copyrighted component of this work in other works.
\normalsize
\maketitle
\thispagestyle{fancy}
\begin{abstract}
Security patterns are intended to support the design and development of secure software systems. However, although established catalogs of security patterns exist, their practical application remains limited. In particular, despite these catalogs, concrete patterns for common security controls such as user authentication (authentication for short) are lacking. This paper aims to make an initial contribution toward closing this gap, as exemplified by authentication. It presents a novel authentication pattern catalog, comprising 14 user authentication patterns. To support the catalog's practical application, it classifies patterns by the well-known concept of authentication factors and by the usual role each pattern fulfills in practice. By cataloging common authentication techniques through authentication patterns, we aim to make an important contribution to supporting software engineers and architects in designing and developing secure software systems.
\end{abstract}
\newtcolorbox{mainBox}[1]{%
		enhanced,
		colback=white!90!black,
		colframe=darkgray,
		colbacktitle=darkgray,
		coltitle=white,
		fonttitle=\bfseries,
		title=#1,
		rounded corners,
		boxrule=3pt,
		drop shadow=darkgray,
		width=\textwidth,
	}

\keywords{Authentication Patterns \and Security Patterns \and User Authentication}

\section{Introduction}\label{sec:intro}

According to the recently updated OWASP Top Ten list, the de facto standard for the most critical security risks in web applications, risks stemming from insecure design remain among the six most frequent categories~\cite{OWASP.Top10}. Yet, designing secure software is both crucial and challenging~\cite{Gressl.2019}. Moreover, due to the shortage of security experts~\cite{Furnell.2020}, non-security experts often handle security-related activities and decision-making~\cite{Gutfleisch.2022,Naji.2025}, underscoring the need to better support engineers and architects in designing secure software systems.

Comprehensive security pattern catalogs exist to support architects and engineers in designing secure software systems~\cite{Fernandez-Buglioni.2013,Schumacher.2006,van.den.Berghe.2022}. However, their adoption in practice remains limited~\cite{Yskout.2015}, due to unclear abstractions in both pattern classification and modeling~\cite{Heyman.2007,van.den.Berghe.2018}. Almost 20 years ago, Heyman et al.~\cite{Heyman.2007} observed that many security patterns describe overly general concepts such as security objectives, guidelines, or principles. This problem persists today, as shown by a recently published catalog~\cite{Cordeiro.2022}. There, broad conceptual categories such as a ``capability'' or an ``asset'' are still listed as security patterns. We argue that this indicates a need for more granular modeling of security patterns to improve practical adoption, and we illustrate this using the well-known security control ``authentication.'' 

\paragraph{Outline:} Section \ref{sec:background} introduces the authentication process and terminology and briefly summarizes our contributions. Section \ref{sec:related} presents related work. The classification is introduced in Section \ref{sec:classification}. Sections \ref{sec:inherence}, \ref{sec:knowledge}, and \ref{sec:possession} then present the authentication patterns in detail. Section \ref{sec:conclusion} concludes the paper, discusses its limitations, and outlines future research directions.

\section{Background and Contributions}\label{sec:background}
\textbf{Authentication} is the process by which a \textbf{subject} verifies the identity they claim to have~\cite{Todorov.2007, NIST.800-63-4}. To verify a claimed identity, a subject uses one or more authenticators. An \textbf{authenticator} is a mechanism or credential that a subject controls and possesses to prove their identity~\cite{Catalog.Paper}. Examples of how authentication can be realized include usernames and passwords for web applications, fingerprint recognition for hardware devices, and one-time passwords (OTPs) sent via SMS or email. We refer to the way in which one or more authenticators are used to perform authentication as an \textbf{authentication technique}.


We note that this paper focuses on authentication techniques for human subjects, i.e., on user authentication, since we have not yet identified consistent authentication patterns for non-human subjects. Thus, for brevity, we will use ``authentication'' to mean ``user authentication'' for the remainder of the paper. Moreover, while ``subject'' is the accurate technical term, we will use the terms ``user'' or ``human'' instead. We will use ``subject`` only when we want to make non-human subjects explicit.

Given the many ways authentication can be realized, the pattern-based nature of recurring authentication techniques to solve recurring authentication problems quickly becomes apparent. However---and even more surprisingly---to our knowledge, no catalog of authentication patterns (for either human or non-human subjects) yet exists. While most established security pattern catalogs include authentication as a security pattern, they limit themselves either by addressing it only as a general concept~\cite{Schumacher.2006, Fernandez-Buglioni.2013, TheOpenGroup, Cordeiro.2022, Distrinet}, or by covering just a few authentication pattern variants~\cite{Distrinet}. Catalogs that cover authentication patterns such as fingerprint-, smart card-, or passkey-based authentication remain missing. A systematic mapping study of authentication patterns that we recently conducted confirmed this gap~\cite{SMS}.

\paragraph{Contribution:} This paper addresses this gap by proposing a catalog of user authentication patterns, comprising 14 distinct authentication patterns. It is the first pattern catalog to recognize authentication as a first-order security concept and to be entirely devoted to patterns related to it. The catalog is primarily intended for non-security experts, such as software architects and engineers. To make the catalog applicable to this audience and ensure a consistent level of abstraction across patterns, we documented them from a solutions architect's perspective, deliberately omitting technical and low-level security details. Furthermore, the catalog organizes patterns by the authentication factor and their usual authentication role to provide initial guidance in the selection process---an essential quality criterion for the usability of security pattern catalogs~\cite{van.den.Berghe.2022}. Table \ref{tab:auth-patterns-primary} provides an overview of the authentication patterns, grouped by their authentication factor at the first level and by their usual authentication role at the second level.

\begin{table*}[!ht]
\scriptsize
\centering
\caption{Overview of the 14 authentication patterns, grouped by their authentication factor and usual authentication role.}
\label{tab:auth-patterns-primary}
\renewcommand{\arraystretch}{1.3}
\begin{tabularx}{\textwidth}{@{\hspace{1pt}}p{0.7cm}| p{0.6cm} | p{2.5cm} X X}
\toprule
\raggedright\textbf{Factor}& \textbf{Usual} \centering\textbf{Role}& \raggedright{\textbf{Authentication} \textbf{Pattern}} & \textbf{Problem} & \textbf{Solution} \\
\midrule

\multirow{4}[28]{*}{\rotatebox[origin=c]{90}{\raisebox{-0.4cm}{\textbf{Inherence-based Authentication}}}}

& \multirow{3}[20]{*}{\rotatebox[origin=c]{90}{\raisebox{-0.35cm}{\textit{Primary}}}}

&
\mbox{\textsc{1. Fingerprint}} \mbox{\textsc{Authentication}}
  & How can a system enable users to identify and authenticate through a physical interaction with a dedicated input device, while minimizing required user interactions?
  
  & Use references of the user's fingerprint(s) as the authenticator. To authenticate, the user presents their fingers to a fingerprint sensor.
  \\

& 
& \mbox{\textsc{2. Face Recognition}} \mbox{\textsc{Authentication}}
  & How can a system enable users to identify and authenticate without physical interaction with a dedicated input device, while minimizing required user effort?
  & Use references of the user's face as the authenticator. To authenticate, the user presents their face to a camera device.
  \\

&
& \mbox{\textsc{3. Iris Recognition}} \mbox{\textsc{Authentication}}
  & How can a system with particularly high security or confidentiality requirements enable users to authenticate themselves within the system while minimizing the risks posed by credential theft?
  & Use references of the user's iris(es) as the authenticator. To authenticate, the user looks into the iris sensor.
  \\ \cmidrule{2-5}

&

\multirow{1}[8]{*}{\rotatebox[origin=c]{90}{\raisebox{-0.35cm}{\textit{Subsidiary}}}}

& \mbox{\textsc{4. Continuous}} \mbox{\textsc{Behavioral-based}} \mbox{\textsc{Authentication}}
&  How can a system continuously verify whether an authentication session is still being operated by its legitimate user, without interrupting the user's workflow?
& Use a behavioral profile of the user's device interactions, learned by the system. Once learned, continuously monitor and compare user interactions with the profile throughout an authentication session.
\\

\midrule


\multirow{5}[38]{*}{\rotatebox[origin=c]{90}{\raisebox{-0.4cm}{\textbf{Knowledge-based Authentication}}}}

& \multirow{2}[12]{*}{\rotatebox[origin=c]{90}{\raisebox{-0.35cm}{\textit{Primary}}}}

& \mbox{\textsc{5. Text Password}} \mbox{\textsc{Authentication}}
  & How can a system authenticate a large population of users accessing personalized content or services, without requiring an underlying PKI infrastructure?
  
  & Use an alphanumeric password only known to the user as the authenticator. To authenticate, the user enters the password.
  \\

& 
& \mbox{\textsc{6. Self-managed}} \mbox{\textsc{PIN Authentication}}
  & How can a system provide fast and convenient authentication through the targeted use of a knowledge-based authenticator?
  & Use a deliberately short or restricted numeric or alphanumeric PIN known to the user as the authenticator. To authenticate, the user enters the PIN.
  
  \\ \cmidrule{2-5}

& \multirow{3}[25]{*}{\rotatebox[origin=c]{90}{\raisebox{-0.35cm}{\textit{Subsidiary}}}}

& \mbox{\textsc{7. Authority-issued}} \mbox{\textsc{PIN Authentication}}
  & How can a system use secret knowledge managed by an issuing authority to further ensure that a user authenticating with an authority-issued authenticator is its rightful owner?
  & Use a short secret PIN issued by the authority through an enrollment process as the authenticator.
  When the user authenticates with the primary authenticator, they are additionally prompted to enter their PIN. \\
  
&
& \mbox{\textsc{8. Dynamic}} \mbox{\textsc{Knowledge-based}} \mbox{\textsc{Authentication}}
  & How can a system further ensure user identity without additional hardware, using personal knowledge that does not require prior user configuration?
  & Use one or more knowledge challenges dynamically derived from the user’s personal data as an additional authenticator. For authentication, the system generates and presents the challenges for the user to answer.
  
  \\

&
& \mbox{\textsc{9. Shared }} \mbox{\textsc{Secrets-based}} \mbox{\textsc{Authentication}}
  & How can a system further ensure a user’s identity without additional hardware, using personal knowledge that the user configures themselves upfront?
  & Use one or more personal knowledge challenges configured by the user as the authenticator. For authentication, the system asks the user one or more of the questions and prompts them to provide the corresponding answers.
  \\
  
\midrule


\multirow{5}[50]{*}{\rotatebox[origin=c]{90}{\raisebox{-0.4cm}{\textbf{Possession-based Authentication}}}}

& \multirow{2}[17]{*}{\rotatebox[origin=c]{90}{\raisebox{-0.35cm}{\textit{Primary}}}}

& \mbox{\textsc{10. Smart Card}} \mbox{\textsc{Authentication}}
  & How can a system enable users to authenticate themselves across multiple devices and access points using a dedicated physical token that is managed by a central authority?
  &  Use a dedicated physical token in the form of a smart card to store the authentication credential.
  To authenticate, the user presents the smart card to a compatible reader.
  
  \\
  
&
& \mbox{\textsc{11. Passkey}} \mbox{\textsc{Authentication}}
  & How can a system authenticate remote users in a phishing-resistant manner without relying on shared secrets and without requiring an infrastructure that uses dedicated physical tokens?
  & Use a platform-internal asymmetric key pair as proof of authentication, stored on the user's device. To authenticate, the system sends a challenge to the user’s device. The device signs the challenge with the private key and returns the signature to the system for verification. 
  
  \\ \cmidrule{2-5}

& \multirow{3}[30]{*}{\rotatebox[origin=c]{90}{\raisebox{-0.35cm}{\textit{Subsidiary}}}}

& \mbox{\textsc{12. Hardware-Token}} \mbox{\textsc{OTP Authentication}}
  &  How can a system provide subsidiary, phishing-resistant authentication using a tamper-resistant approach that operates independently of the user’s general-purpose devices?
  &  Use a dedicated, physical hardware token that generates a dynamic, short-lived one-time password (OTP). To authenticate, the user uses the token to generate an OTP and transmit it to the system for verification. 
  
  \\

&
& \mbox{\textsc{13. Software-Token}} \mbox{\textsc{OTP Authentication}}
  & How can a system provide subsidiary authentication that is applicable to a broad user base by using the user’s personal device to generate authentication credentials?
  & Use a software-based authentication application installed on the user’s personal device to generate OTPs. To authenticate, the user opens the authentication application, which generates an OTP from the shared secret and transmits it to the system for verification. \\

&
& \mbox{\textsc{14. Out-of-Band}} \mbox{\textsc{OTP Authentication}}
  & How can a system provide subsidiary authentication to a broad, potentially unregistered user base by using an out-of-band communication channel the user already has access to?
  & Let the user configure one or more out-of-band communication channels in a dedicated registration step. To authenticate, the system generates an OTP, stores it for verification, and transmits it via one of the channels. The user retrieves the OTP from their out-of-band channel and submits it to the system. \\ 

  \bottomrule

\end{tabularx}
\end{table*}

\section{Related Work}\label{sec:related}

This pattern catalog is related to established security pattern catalogs. Yoder and Barcalow~\cite{Yoder.1997} laid the groundwork for security patterns in 1997. Since then, multiple established security pattern catalogs have emerged. The most established catalogs are those of Schumacher et al.~\cite{Schumacher.2006} and Fernandez-Buglioni~\cite{Fernandez-Buglioni.2013}. Among others, both introduce patterns for identity management and for authentication, including the ``Authenticator'' pattern~\cite{Schumacher.2006, Fernandez-Buglioni.2013}. However, both catalogs remain highly abstract, as evidenced in a later work by Fernandez et al.~\cite{Fernandez.2022}, which reclassifies the Authenticator pattern as an abstract pattern. In addition, the ``Security Design Patterns'' catalog by The Open Group addresses authentication as a pattern but does not document it in greater detail~\cite{TheOpenGroup}. The catalog developed by Cordeiro et al. contains some authentication-related patterns, such as the aforementioned Authenticator pattern~\cite{Cordeiro.2022}. However, as a systematic mapping study, it is limited to listing patterns already proposed in related work; it does not include new authentication-related patterns. The only security pattern catalog we know of that documents authentication-related patterns in more detail---albeit limited---is proposed by the Distrinet group~\cite{Distrinet, van.den.Berghe.2022}. Their catalog lists the abstract ``Authentication'' pattern, along with four implementations: ``Password-based authentication'' and three token-based authentication patterns. None of these catalogs covers authentication patterns in sufficient detail to document the recurring problems, solutions, and set of forces of the various authentication alternatives, limiting their effective use in practice.

Moreover, our previously published catalog of authentication techniques is closely related to this paper. At the time of writing, it lists 39 different authentication techniques. As introduced above, an authentication technique refers to the conceptual way in which one or more authenticators are used to perform authentication. As such, they are closely related, but not equivalent to authentication patterns. Both catalogs support architects and engineers in different ways to design secure software systems: authentication techniques as concrete instances of how authentication is conceptually realized; authentication patterns as guidelines for designing authentication techniques. There are two ways in which an authentication technique can be associated with an authentication pattern: 1.) A technique or its application may exhibit pattern characteristics, e.g., by representing a systematic, repeatedly applied conceptual approach to performing authentication in a practical context, as is the case with password-based authentication; 2.) Techniques may exhibit cross-technique pattern characteristics, e.g., when different biometric features are implemented similarly to achieve passive user authentication. To create the authentication pattern catalog, we analyzed the authentication techniques based on these two characteristics and assessed whether they solve recurring problems under a set of forces~\cite{Catalog.Paper}.

\section{Classification of Authentication Patterns} \label{sec:classification}

Security patterns can be classified based on numerous characteristics, such as the development phase in which they are applied and the security objectives they are intended to achieve~\cite{Cordeiro.2022, Yskout.2007}. Authentication patterns share similar properties~\cite{Catalog.Paper}. We followed the same approach and classified authentication patterns based on two properties: the authentication factor inherent to the authenticator employed by a pattern and the role that the pattern usually plays in the context of authentication. Both properties provide important guidance for navigating the catalog and selecting an appropriate authentication pattern for a specific context, thereby fulfilling a crucial requirement for a pattern catalog in the security context: providing selection guidance~\cite{van.den.Berghe.2022}.

\subsection{Authentication Factor}
The authentication factor is an established classification approach for authenticators~\cite{NIST.800-63-4}. It is most commonly divided into three factors, which is also reflected in the authentication pattern catalog:

\begin{itemize}[leftmargin=0pt]
    \item[] \textbf{Inherence-based:} The authenticator is based on an inherent feature of the user. For humans, these are typically biometric features.
    \item[] \textbf{Knowledge-based}: The authenticator is based on secret information that the user has.
    \item[] \textbf{Possession-based:} The authenticator is based on something the user possesses.
\end{itemize}

The authentication pattern catalog organizes patterns at the first level by the authentication factor. We decided this for two reasons. First, the authentication factor is arguably the most commonly used approach for classifying authentication techniques in practice. By incorporating this classification into the pattern catalog's structure, we aim to make it easier for practitioners to access patterns, given the concept's widespread recognition.

Second, each factor is characterized by a set of properties that are inherent to the patterns of that factor. For example, inherence-based authentication patterns inherently involve using and processing highly sensitive data and are therefore subject to particularly strict data protection requirements. By grouping patterns by the authentication factor, the catalog highlights the relationship between a pattern's factor-inherent and pattern-specific properties, particularly when comparing a pattern for a specific factor with alternative patterns for the same factor.

\subsection{Usual Authentication Role}
Although an authentication pattern does not inherently restrict the role it plays in the authentication context, all patterns we identified are usually implemented with a specific role. This usual role provides an important criterion for selecting an authentication pattern and therefore constitutes the second characteristic by which the catalog classifies the patterns. Two usual authentication roles can be distinguished: patterns for primary authentication and patterns for subsidiary authentication.

\begin{itemize}[leftmargin=0pt]
    \item[] \textbf{Primary:} An authentication pattern is a primary authentication pattern if it is usually implemented as the central entry point for accessing restricted resources or services of a software system. 
    \item[] \textbf{Subsidiary:} An authentication pattern is a subsidiary authentication pattern if it is usually implemented to support or enhance a primary authentication pattern, or to authenticate users in use cases that are not a central entry point into a software system.
\end{itemize}

The concrete subsidiary role an authentication pattern can have varies by use case. This category includes patterns usually used as a second factor in multi-factor authentication, as well as patterns typically implemented as fallback authentication techniques or used in cases not primarily intended to provide access to central, protected services or resources. 

We illustrate both roles using text password authentication and security questions, as are typically used in web applications. Both patterns implement user authentication but differ in their usual role. Passwords usually let users authenticate so they can use the application's services. Security questions authenticate users who forgot their password. Thus, passwords usually play a more central role in authentication than security questions, which play a more subsidiary role. Without a functioning password authentication mechanism, users cannot use a web application's services at all. With password authentication but without security questions, users can still use the application’s services, though usability is significantly limited if they forget their password. 

\subsection{Using the Catalog} Table~\ref{tab:auth-patterns-primary} provides an overview of the authentication pattern catalog. For each pattern, it provides a short description of the pattern's underlying problem and how it solves it. The table follows the same structure as the catalog: at the first level, it organizes the authentication patterns by the authentication factor. Then, for each authentication factor, it organizes the corresponding patterns at the second level by their usual role, beginning with the primary patterns and ending with the subsidiary patterns. 

Next, we present all 14 authentication patterns. Their presentation follows the same order as in Table~\ref{tab:auth-patterns-primary}: section-wise, we first introduce the class-level properties of each authentication factor, then describe each authentication pattern in detail in terms of its context, the problem it addresses, the underlying forces, the pattern's solution, including the consequences of using the pattern, and known uses of the pattern.



\section{Inherence-based Authentication Patterns}\label{sec:inherence}

Inherence-based authentication patterns offer highly appealing authentication features. They typically involve biometric data. As a result, they enable the authenticator to both identify and authenticate humans simultaneously, thereby offering a particularly high level of usability and comfort in authentication~\cite{Todorov.2007, Maltoni.2022}. As a result, they offer alternatives to possession-based authentication patterns for reliably and quickly authenticating users without incurring the overhead and costs of issuing, distributing, and maintaining user-specific authenticators or public-key infrastructures (PKI).

However, this feature also comes at a price. As all inherence-based authentication patterns we know of involve biometric data, they are infeasible to authenticate non-human subjects\footnote{At the time of writing, there are no patterns we know of. However, in the era of AI, we will not be surprised if such patterns evolve in the near future.}. Moreover, because biometric data is highly sensitive, such patterns require strong data security and privacy~\cite{Langenderfer.2005}. Consequently, they are primarily suitable for local authentication, that is, to authenticate a user on a local device. Using inherence-based authentication patterns for remote authentication raises serious security and privacy concerns. Therefore, these patterns should generally never be used for remote authentication~\cite{NIST.800-63-4}. However, if the operational context requires and the legal situation allows the central storage of biometric references, they must be stored in a dedicated, access-controlled repository. Transmission must be protected by strong encryption, and the stored references must be cryptographically protected at rest. In such deployments, particular care must be taken to comply with applicable data protection regulations~\cite{Langenderfer.2005}. 

Moreover, unlike other authentication factors, inherence-based authentication patterns can be inaccurate. This is reflected in false-positive and false-negative rates, i.e., even when a valid authenticator is presented, the pattern can still deem it invalid~\cite{Alwahaishi.2020}. The inaccuracy requires strict lockout policies and fallback authentication patterns when using them. Moreover, heavy reliance on specialized sensor hardware exacerbates this issue, as it is susceptible to environmental conditions, failures, and damage~\cite{Vielhauer.2010}. Lastly, specialized sensor hardware adds acquisition and maintenance costs.

\subsection{Primary Authentication}
We identified three primary inherence-based authentication patterns: fingerprint, face recognition, and iris recognition. \pagebreak

\pattern{Fingerprint Authentication}

\noindent \textbf{Context:} A system is designed for use in an environment where users are expected to identify and authenticate themselves frequently and in short time intervals, where the solution can rely on users physically interacting with a dedicated input device, and where the overhead and costs for issuing, distributing, and maintaining specially issued, dedicated, user-specific authenticators are unreasonable. To meet this requirement, a solution is needed that enables users to identify and authenticate themselves to the system with minimal interaction with the input device.

\noindent \textbf{Problem:} How can a system enable users to identify and authenticate themselves through physical interaction with a dedicated input device while minimizing user interaction and without relying on specially issued, dedicated, user-specific authenticators? 

\noindent \textbf{Forces:} 
\begin{itemize}[leftmargin=1em, topsep=0pt]
    \item[-] \textit{Minimal interaction vs. Sample completeness:} Keeping user interactions with the input device to a minimum encourages the use of the input device for as short a time as possible; however, the data collected must still be complete and stable enough to allow for a reliable comparison.
    \item[-] \textit{Convenience vs. Environmental reliability:} A brief, effortless physical interaction is convenient and quick; however, it is the reliance on direct physical contact that makes the capture process vulnerable to the condition of the contact surface, which can be affected by environmental factors.
    \item[-] \textit{Residues from physical contact:} Physical contact with a surface can leave behind residues that could be recovered and reused later, which can pose particular threats.
\end{itemize}

\noindent \textbf{Solution:} Use a dedicated fingerprint sensor to capture one or more samples of the user's fingerprint. In a dedicated registration step, the system extracts a reference from these samples and stores it in a secure environment, usually on the authenticating local device. When authentication is requested, the user presents their finger to the sensor, which captures a new sample and compares it against the stored reference. If the comparison yields sufficient similarity, the user is authenticated and granted access~\cite{Maltoni.2022}.

To prevent unauthorized access, the system should particularly protect against fingerprint presentation attacks~\cite{Gonzalez-Soler.2021}. 

\noindent \textbf{Consequences:}

\begin{itemize}[leftmargin=0pt, topsep=0pt]
    \item[] \textit{Benefits:}
    \begin{itemize}[topsep=0pt]
        \item Authentication requires only a single, brief physical gesture, minimizing user effort and interaction time.
        \item Mobile devices are operated with the hands anyway, making this pattern particularly attractive.
        \item In everyday settings, fingerprint sensors offer very high authentication accuracy.
    \end{itemize}
    \item[] \textit{Liabilities:}
    \begin{itemize}[topsep=0pt]
        \item The pattern is highly vulnerable to presentation attacks involving latent fingerprints left on surfaces, requiring active countermeasures to maintain authentication integrity.
        \item Environmental conditions, such as wet, dirty, or injured fingers, can compromise authentication accuracy, thereby reducing the pattern's usability and even suitability for authentication.
        \item As this pattern involves deliberate touch gestures, it is not suitable for authenticating certain user groups, including people with specific skin conditions, worn-out fingerprint ridges, or physical impairments affecting their fingers or hands.
    \end{itemize}
\end{itemize}

\noindent \textbf{Known Uses:} Fingerprint authentication is well-established on most mobile devices~\cite{Bhagavatula.2015, Kim.2021}. Many mobile apps with high security requirements, such as mobile banking~\cite{Sharma.2018} and payment applications~\cite{Jo.2016}. Automated border control and e-Passports~\cite{Labati.2017}.  \newpage


\pattern{Face Recognition Authentication}

\noindent \textbf{Context:} A system is designed for use in a context where users frequently need to identify and authenticate themselves within the system, but where physical interaction with a dedicated input device is not possible, required, or even prohibited. To meet this requirement, a solution is needed that enables users to identify and authenticate themselves to the system with minimal effort, using alternative input devices that can be operated without physical contact.

\noindent \textbf{Problem:} How can a system enable users to identify and authenticate themselves without physical interaction with a dedicated input device, while minimizing the required user effort?

\noindent \textbf{Forces:} 
\begin{itemize}[leftmargin=1em, topsep=0pt]
    \item[-] \textit{Minimal interaction vs. User positioning:} Minimizing the effort required of the user calls for requiring as few conscious actions as possible; however, reliable contactless data collection can depend on the user's correct proximity, angle, and orientation.
    \item[-] \textit{Passive authentication vs. Live-presence assurance:} Authentication that does not require any conscious action on the part of the user essentially enables passive user authentication; however, the lack of active confirmation poses the challenge for the system to ensure that it is interacting with a user who is actually present and alive.
    \item[-] \textit{Data sensitivity and privacy concerns:} Contactless authentication using alternative input devices carries the risk of unintentionally capturing unwanted, sensitive data and data of other users in addition to the user being authenticated, which can pose privacy and confidentiality risks.
\end{itemize}

\noindent \textbf{Solution:} Use a camera device to capture one or more images of the user’s face. In a dedicated registration step, the system extracts a reference from these images and stores it in a secure environment, usually on the authenticating local device. For authentication, the user positions themselves in front of the camera, which captures a new image and compares it to the stored reference. If the comparison yields a sufficient degree of similarity, the user is authenticated and granted access~\cite{Chowdhury.2017, Li.2005}.

To prevent unauthorized access, the system should include liveness detection to distinguish a physically present user from a photograph or video playback~\cite{Chakraborty.2014}.

\noindent \textbf{Consequences:}

\begin{itemize}[leftmargin=0pt, topsep=0pt]
    \item[] \textit{Benefits:}
    \begin{itemize}[topsep=0pt]
        \item Authentication does not require physical contact with a special device, making the process suitable for situations where hygiene is a concern and for people who have difficulty using fingerprint sensors.
        \item Since no deliberate interaction with a device is required, this pattern is also suitable for passive user authentication.
        \item The pattern relies solely on camera hardware, which is widely used in consumer devices, thereby offering broad applicability.
    \end{itemize}
    \item[] \textit{Liabilities:}
    \begin{itemize}[topsep=0pt]
        \item Facial appearance can change due to facial hair, cosmetics, or injuries, potentially reducing the pattern's usability.
        \item Environmental conditions, such as dark environments and poor lighting, can compromise authentication accuracy.
        \item Data collection with a camera poses the risk that data from bystanders may be accidentally captured, raising privacy concerns and legal issues.
    \end{itemize}
\end{itemize}

\noindent \textbf{Known Uses:} Many iOS, Android, and Windows mobile consumer devices offer face recognition authentication for local authentication~\cite{Bertok.2016}. Automated border control and e-Passports~\cite{Labati.2017}. Solutions for physical access control~\cite{Taleb.2014}, such as Alcatraz AI~\cite{alcatrazai}. \newpage

\pattern{Iris Recognition Authentication}

\noindent \textbf{Context:} A system is designed for use in an environment where particularly high security or confidentiality requirements apply and where users must identify and authenticate themselves within the system. To meet this requirement and protect the authentication process from authentication data theft, a solution is needed that enables users to identify and authenticate themselves as securely and reliably as possible using biometric data.

\noindent \textbf{Problem:} How can a system with particularly high security or confidentiality requirements enable users to authenticate themselves within the system while minimizing the risks posed by credential theft?

\noindent \textbf{Forces:} 
\begin{itemize}[leftmargin=1em, topsep=0pt]
    \item[-] \textit{High matching precision vs. Required hardware:} The high-security context calls for a biometric authenticator that offers very high matching accuracy and long-term stability; however, achieving this level of accuracy typically requires more specialized and expensive capture hardware than less accurate alternatives.
    \item[-] \textit{High security vs. Throughput:} Meeting high security requirements may necessitate deliberate and precise interaction with the capture device, for example, to minimize the false-positive rate; however, precise interaction reduces throughput and makes the solution less suitable for scenarios involving moving or impatient users, as well as high access volumes.
    \item[-] \textit{Inherent Resistance to Forgery:} Given the high security requirements, the underlying biometric authenticator's inherent resistance to forgery and counterfeiting becomes a key distinguishing factor among biometric options.
\end{itemize}

\noindent \textbf{Solution:} Use a dedicated iris recognition sensor to capture one or more samples of the user's iris. In a dedicated registration step, the system extracts a reference from these samples and stores it securely. For authentication, the user stands in front of the sensor, which captures a new image and compares it with the stored reference. If the comparison yields sufficient similarity, the user is authenticated and granted access~\cite{Bowyer.2016}.

To prevent unauthorized access, the system should include liveness detection to distinguish a physically present user from a photograph or artificial eye~\cite{Chakraborty.2014}.

\noindent \textbf{Consequences:}

\begin{itemize}[leftmargin=0pt, topsep=0pt]
    \item[] \textit{Benefits:}
    \begin{itemize}[topsep=0pt]
        \item The iris is highly unique and remains stable throughout a person's entire life, ensuring high accuracy and very low false positive rates compared to alternative inherence-based patterns.
        \item The iris provides high security as it is significantly more resistant to forgery than other biometric features.
    \end{itemize}
    \item[] \textit{Liabilities:}
    \begin{itemize}[topsep=0pt]
        \item Requires specialized, expensive sensor hardware, limiting widespread use compared to alternative inherence-based patterns.
        \item Requires deliberate, precise user positioning in front of the sensor, making it less suitable for contexts where users are moving or throughput is a priority.
        \item Environmental conditions such as poor lighting, reflections on glasses, or excessive distance from the sensor can compromise authentication accuracy.
    \end{itemize}
\end{itemize}

\noindent \textbf{Known Uses:} India's Aadhaar program~\cite{Bowyer.2016}. UAE border control~\cite{Rathgeb.2012}. CERN access control for particle physics research facilities~\cite{Murad.2021}. Apple's Vision Pro and Microsoft's HoloLens 2~\cite{Wang.2025}. 

\addvspace{30pt}

\subsection{Subsidiary Authentication} We identified one subsidiary inheritance-based pattern: continuous behavioral-based authentication. 

\addvspace{10pt}

\pattern{Continuous Behavioral-based Authentication}

\noindent \textbf{Context:} A system is designed for use in an environment where an authentication session may extend over a long period of time, creating the risk that the session could be taken over by an unauthorized person after the initial authentication. To mitigate this risk, a solution is needed that continuously verifies whether the authentication session is still being maintained by the legitimate user, without interrupting the user's workflow.

\noindent \textbf{Problem:} How can a system continuously verify whether an authentication session is still being operated by its legitimate user, without interrupting the user's workflow?

\noindent \textbf{Forces:} 
\begin{itemize}[leftmargin=1em, topsep=0pt]
    \item[-] \textit{Detection sensitivity vs. Unintended session disruption:} As the user's workflow should not be interrupted, observing user-specific features or behavior requires some detection sensitivity. If it is too weak, an actual session takeover will go undetected; if it is too strict, the user's workflow will be interrupted, as the legitimate user's session might be terminated or interrupted.
    \item[-] \textit{Behavioral instability vs. Profile stability:} Authentication requires a stable, reliable profile of the user's behavior. However, user behavior may change over time. While the profile can be designed to adapt dynamically to such changes, this raises the question of how the system can distinguish legitimate, gradual changes from a gradual, intentional attempt to deceive.  
    \item[-] \textit{Continuous observation vs. Privacy:} To verify a user’s identity throughout an entire session, the system must continuously monitor and record the user’s interactions over the entire duration of that session. This raises serious privacy concerns.
\end{itemize}

\noindent \textbf{Solution:} Let the system learn how each user interacts with it by analyzing multiple behavioral characteristics and recording these interactions in a behavioral profile. During an initial training phase, the system records the behavioral characteristics while the user interacts with it and compiles them into an initial behavioral profile. After the training phase, the system continuously monitors the user’s behavioral characteristics throughout the entire authentication session and compares them to the stored profile. If the observed behavior shows sufficient similarity to the stored profile, the session continues without interruption.  Otherwise, the authentication session will be terminated~\cite{Fitzgerald.2018}.

To prevent unauthorized access, the system should set a threshold below which the similarity score triggers additional authentication or terminates the session~\cite{TypingDNA}.

\noindent \textbf{Consequences:}

\begin{itemize}[leftmargin=0pt, topsep=0pt]
    \item[] \textit{Benefits:}
    \begin{itemize}[topsep=0pt]
        \item The pattern continuously verifies the user's identity during an authentication session, thereby detecting attempts to take over the session that active authentication patterns cannot detect after initial authentication.
        \item Authentication is performed entirely passively from the user's perspective, so the user's workflow is not interrupted.
    \end{itemize}
    \item[] \textit{Liabilities:}
    \begin{itemize}[topsep=0pt]
        \item The authentication accuracy depends entirely on the quality of the behavioral profile created during the training phase.
        \item User behavior can vary over time and in response to external factors, such as injuries, stress, or changes in interaction habits, which can ultimately affect accuracy.
        \item Monitoring user behavior raises significant privacy concerns, as the system must continuously observe and record users' interactions throughout the entire session.
    \end{itemize}
\end{itemize}

\noindent \textbf{Known Uses:} TypingDNA's ActiveLock\cite{TypingDNA} employs it to verify that a device is being operated by its legitimate user. BioCatch~\cite{biocatch} and LexisNexis~\cite{behaviosec} offer solutions to detect fraud during authenticated sessions; Aetna~\cite{aetna} uses it for the same reason in healthcare. Online proctoring platforms, such as Meazure Learning~\cite{meazurelearning} and Pearson VUE~\cite{pearsonvue}, apply it to verify student identities throughout examinations. \newpage

\section{Knowledge-based Authentication Patterns}\label{sec:knowledge}

Knowledge-based authentication patterns form the second class of authentication patterns. In contrast to inherence-based authentication patterns, which bind the subject to the authenticator used and are inherently non-anonymous, knowledge-based patterns separate subject identification from authentication. Consequently, authentication data derived from knowledge-based patterns is generally less sensitive than that derived from inherence-based ones. This enables anonymous or pseudonymous authentication of subjects, which offers significant advantages by enabling authentication of non-human subjects, transmitting authentication data to remote systems, and storing it in central repositories; all of which are significant limitations that apply to inherence-based patterns. Furthermore, knowledge-based authentication patterns can be used universally and are inherently location-independent, since they do not rely on either physical tokens issued in advance or an existing PKI infrastructure. This is also why they are more cost-effective to deploy than inherence-based and possession-based patterns.  


However, knowledge-based patterns also have significant disadvantages compared to inherent and possession-based ones. First, the universal applicability of these patterns comes at the expense of memorability, which severely limits their usability~\cite{Bonneau.2012}. This fact also requires recovery mechanisms, including the corresponding processes and infrastructure, for events in which a subject's secret knowledge is completely lost. In addition, they raise significant security concerns, as they can be easily stolen and are therefore frequently targeted by fraud attempts, such as phishing attacks~\cite{Bonneau.2012}. Consequently, such patterns are often considered insufficient and should not be used as the sole authentication factor whenever security is a concern~\cite{NIST.800-63-4}.

\subsection{Primary Authentication} \sloppy{We identified two primary knowledge-based authentication patterns: text password and self-managed PIN authentication.} 

\addvspace{10pt}

\pattern{Text Password Authentication}

\noindent \textbf{Context:} A system is designed to enable a potentially large and diverse population of users to access personalized content or services while minimizing the sensitivity of the personal data transmitted in the process. To meet this requirement, a solution is needed that is universally applicable, does not require the prior issuance of physical access credentials, and can be implemented without specialized hardware on either side of the authentication process.

\noindent \textbf{Problem:} How can a system authenticate a potentially large and diverse population of users accessing personalized content or services, without requiring an underlying PKI infrastructure?

\noindent \textbf{Forces:} 
\begin{itemize}[leftmargin=1em, topsep=0pt]
    \item[-] \textit{High familiarity vs. Low security awareness:} Widespread familiarity with using a secret for authentication lowers the barriers to adoption and eliminates the need for user training; however, it also reinforces habits, such as using the same secret across different systems, that cannot be fully mitigated by policies alone.
    \item[-] \textit{Secret memorability vs. Secret strength:} The shorter and less complex the secret used is, the more memorizable it is for the user; however, this conflicts with the secret strength and, thus, with the level of security the secret provides.
    \item[-] \textit{User self-service vs. Security dependency:} Self-managed enrollment and management of user credentials improve the usability and accessibility of the authentication technique and enable cost-effective, large-scale deployment; however, the resulting minimum level of security provided by the credentials depends more on the user than on the system’s guarantees, and any policy restrictions the system imposes to raise this minimum level result in friction that directly impairs usability and accessibility.
    \item[-] \textit{Recovery self-service vs. Additional attack vectors:} Secrets can be forgotten, which is why a recovery self-service is essential for the usability of the authentication technique; however, this leads to additional authentication channels and, consequently, additional attack vectors.
\end{itemize}

\noindent \textbf{Solution:} Allow users to define credentials, including their user identifier and a secret text string (the password), during a dedicated registration step. The system stores the credentials in hashed form. To authenticate, the user submits their credentials, which the system compares with the stored hash. If the submitted secret matches the stored reference, the system authenticates the user and grants access~\cite{Conklin.2004}.

To prevent unauthorized access, the system should enforce strict password constraints, such as requiring high password complexity, preventing users from using commonly used or previously compromised passwords, and enforcing a lockout policy after a certain number of failed authentication attempts.

\noindent \textbf{Consequences:}

\begin{itemize}[leftmargin=0pt, topsep=0pt]
    \item[] \textit{Benefits:}
    \begin{itemize}[topsep=0pt]
        \item Passwords offer an easily accessible authentication technique for everyday software systems; most users are familiar with them and can use them without training or instruction, and the effort required to register login credentials is typically modest.
        \item The technique does not require any special hardware beyond a standard keyboard or touchpad, enabling a wide range of applications and use cases.
        \item Since there are no specific hardware or infrastructure requirements, the costs of implementing and operating text password authentication are low compared to alternative approaches.
    \end{itemize}
    \item[] \textit{Liabilities:}
    \begin{itemize}[topsep=0pt]
        \item Apart from password strength guidelines, the system has no influence on how secure a password is, since it is chosen by the user. 
        \item Users must memorize their passwords. This imposes a cognitive burden and requires password-recovery mechanisms when users forget their login credentials, introducing additional attack vectors. 
        \item Users tend to use the same login credentials across different applications, which poses additional security risks.
        \item Passwords are often targets of security attacks, such as phishing, credential stuffing, and brute-force attacks.
    \end{itemize}
\end{itemize}

\noindent \textbf{Known Uses:} Virtually every web application, operating system, and network service~\cite{Kumar.2023}. \newpage 

\pattern{Self-managed PIN Authentication}

\noindent \textbf{Context:} A system is designed for use in an environment where users must frequently authenticate themselves and where, therefore, a particularly user-friendly authentication technique is desired. Because repeatedly entering a complex password would be highly inconvenient for users, and because inherence-based authentication patterns may not always be available due to hardware limitations, a solution is needed that lets users authenticate quickly using a simple, self-chosen secret.

\noindent \textbf{Problem:} How can a system provide fast and convenient authentication through the targeted use of a knowledge-based authenticator?

\noindent \textbf{Forces:} 
\begin{itemize}[leftmargin=1em, topsep=0pt]
    \item[-] \textit{Authentication convenience vs. Credential strength:} Deliberately limiting the length and complexity of the secret minimizes the effort required for memorization and entry, thereby enabling fast and frequent authentication; however, this limitation directly weakens the system’s resilience against guessing and brute-force attacks, meaning that the practical viability of the method depends entirely on compensating countermeasures.
    \item[-] \textit{Countermeasure completeness vs. Pattern viability:} Since the deliberately low level of complexity of the secret itself offers no significant protection, the effectiveness of the pattern depends entirely on a comprehensive set of strict countermeasures; omitting or weakening a single countermeasure can therefore render the pattern useless as a security measure.
    \item[-] \textit{Authentication Scope vs. Applicability:} Due to the risks posed by a deliberately easy secret, its use should be limited to local authentication in order to mitigate these risks; however, this limitation implies that the pattern is not applicable as a standalone solution for systems that require remote authentication, requiring the system to provide supplementary authentication techniques for these scenarios.
\end{itemize}

\noindent \textbf{Solution:} Use text password authentication with a deliberately short numeric or alphanumeric password (the PIN). Mitigate the risks of the PIN's intentionally low complexity with strict countermeasures. These include preventing users from choosing commonly used or easily guessable PINs, enforcing a lockout policy after a limited number of failed authentication attempts, and restricting the use of this pattern to local authentication on end devices.

\noindent \textbf{Consequences:}

\begin{itemize}[leftmargin=0pt, topsep=0pt]
    \item[] \textit{Benefits:}
    \begin{itemize}[topsep=0pt]
        \item A short, simple password is much easier to remember than a complex one, which reduces the cognitive burden on users.
        \item Entering a PIN requires little effort on the user's part, allowing even frequent authentication to be carried out quickly and conveniently.
        \item Self-managed PIN authentication requires only a keyboard device for input and can therefore be implemented on virtually any end device.
    \end{itemize}
    \item[] \textit{Liabilities:}
    \begin{itemize}[topsep=0pt]
        \item PINs are very simple and are therefore highly vulnerable to guessing and brute-force attacks.
        \item Due to their high vulnerability, their effectiveness depends entirely on the implementation of appropriate mitigation strategies.
        \item People often use mnemonics, such as birthdays, which makes them particularly vulnerable to social engineering attacks.
        \item As PINs ensure only a very low level of security, self-managed PIN authentication is unsuitable for remote authentication.
    \end{itemize}
\end{itemize}

\noindent \textbf{Known Uses:} Virtually every mobile device, such as smartphones, tablets, and laptops~\cite{Zezschwitz.2013}. Commonly used as alternatives to inherence-based authentication patterns for mobile apps with high security requirements, such as mobile banking apps~\cite{Kruzikova.2022}.

\addvspace{30pt}

\subsection{Subsidiary Authentication} We identified three knowledge-based authentication patterns: authority-issued PIN, dynamic knowledge-based, and shared secrets-based authentication. \newpage

\pattern{Authority-issued PIN Authentication}

\noindent \textbf{Context:} A system is designed for use in an environment where the authenticator used for primary authentication is issued and managed by a central authority, and where the presentation of the authenticator alone is considered insufficient to verify that the user presenting it is its rightful owner. To meet this requirement, a solution is needed that allows the central authority to link the authenticator to its rightful owner via an additional authenticator.

\noindent \textbf{Problem:} How can a system use an additional authenticator managed by an issuing authority to further ensure that a user authenticating with an authority-issued authenticator is its rightful owner?

\noindent \textbf{Forces:} 
\begin{itemize}[leftmargin=1em, topsep=0pt]
    \item[-] \textit{Out-of-Band Security vs. Availability Delay:} Providing the login credentials via a channel separate from the primary authenticator reduces the risk of both authenticators being compromised simultaneously; however, the physical or procedural nature of this provision can lead to delays in the initial availability of the authenticator, limiting usability and quick accessibility.
    \item[-] \textit{Authority-controlled secret vs. Memorization burden:} By having the issuing authority assign login credentials at the time of issuance, the risk of weak or easily guessed secrets is eliminated; however, it increases memorization burden on the part of the user,  as they must memorize a secret with which they have no personal connection.
    \item[-] \textit{Secret changeability vs. Security guarantees:} To enhance a secret's memorability, users can be allowed to change the secret upon its issuance; while this increases usability, it undermines the security guarantees provided by the issuance through the authority.
    \item[-] \textit{Authority-managed secret lifecycle vs. Administrative overhead:} Centralizing the issuance, suspension, and reissuance of credentials under a single authority enables systematic lifecycle management of access credentials and reduces the effort required for recovery in the event of a compromise; however, this centralization requires administrative processes and infrastructure that entail additional effort compared to self-service alternatives.
\end{itemize}

\noindent \textbf{Solution:} Use a short secret PIN assigned to the user by the issuing authority as an additional knowledge-based authenticator. During an enrollment process managed by the issuing authority, the authority assigns a PIN to the user, registers it in the system, and delivers it to the user through a secure out-of-band channel. To authenticate, the user presents the authority-issued authenticator, upon which the system prompts the user to enter the PIN. If the transmitted PIN matches the stored reference, the user is authenticated and granted access~\cite{Renaud.2015}.

To prevent unauthorized access, the system should set a validity period for the PIN's first use and apply a lockout policy after a limited number of failed authentication attempts.

\noindent \textbf{Consequences:}

\begin{itemize}[leftmargin=0pt, topsep=0pt]
    \item[] \textit{Benefits:}
    \begin{itemize}[topsep=0pt]
        \item Out-of-band delivery provides additional security as it reduces the risk of the PIN and the primary authenticator being compromised simultaneously.
        \item Authority control over the PIN enables a centralized revocation and reissuance process in the event of a compromise.
        \item Since the PIN is initially assigned by the issuing authority rather than chosen by the user, the risk of weak or easily guessable PINs is eliminated at the time of issuance.
    \end{itemize}
    \item[] \textit{Liabilities:}
    \begin{itemize}[topsep=0pt]
        \item It is common practice for users to be able to change their PIN after initial use. However, this negates the initial benefit that the authority-issued PIN is not weak or easy to guess.
        \item The registration process administered by the issuing authority entails additional administrative burdens compared to self-administered alternatives.
        \item Delivery via the out-of-band channel can take several days, severely limiting the pattern's usability on its first use.
    \end{itemize}
\end{itemize}

\noindent \textbf{Known Uses:} EMV cards in retail banking at point-of-sale terminals and ATMs~\cite{Renaud.2015}. U.S. government smart cards, including the Common Access Card (CAC) for Department of Defense employees~\cite{SmartCardAlliance} and the Personal Identity Verification (PIV) card for civilian federal agencies~\cite{PIV}. Mobile network operators issue PINs for their SIM cards to the SIM card holders~\cite{Chakraborty.2019}. \newpage

\pattern{Dynamic Knowledge-based Authentication}

\noindent \textbf{Context:} A system is designed for use in an environment where subsidiary authentication is required for a primary authentication pattern, but where neither specialized hardware nor prior configuration by the user is available or practical. To meet this requirement, a solution is needed that enables the system to verify a user’s identity using personal knowledge unique to the user, without requiring prior registration of that knowledge.

\noindent \textbf{Problem:} How can a system further ensure user identity without additional hardware, using personal knowledge that does not require prior user configuration?

\noindent \textbf{Forces:} 
\begin{itemize}[leftmargin=1em, topsep=0pt]
    \item[-] \textit{Configuration-free deployment vs. Data source dependency:} The independence from a prior configuration allows a deployment in a wide range of use cases; however, the challenge data depends on the availability and scope of external or historical data sources, meaning that authentication cannot be performed for users for whom such data is not accessible.
    \item[-] \textit{Challenge unpredictability vs. User recall:} Generating challenges dynamically from personal data prevents users from preparing or sharing answers in advance, reducing the risk of sharing the secrets; however, this also carries the risk that a user may not remember the correct answer, undermining authentication accuracy and usability in ways the system cannot anticipate.
    \item[-] \textit{External data utility vs. Privacy exposure:} Since no prior configuration is required, it can be necessary to access third-party or historical user data; however, this may raise serious privacy concerns and be subject to legal or regulatory restrictions that limit the pattern's applicability.
\end{itemize}

\noindent \textbf{Solution:} Use one or more knowledge challenges derived from the user's personal data as an additional authenticator. The system constructs challenges at the time of authentication using historical or external data, without requiring prior user configuration. To authenticate, the system presents the challenges to the user, who provides their answers. If the answers provided match the expected values derived from the data sources, the user is authenticated and granted access.

To prevent unauthorized access, the system should limit the number of challenges presented per authentication attempt and apply a lockout policy after a limited number of failed attempts.

\noindent \textbf{Consequences:}

\begin{itemize}[leftmargin=0pt, topsep=0pt]
    \item[] \textit{Benefits:}
    \begin{itemize}[topsep=0pt]
        \item The pattern requires no prior user configuration, enabling deployment without onboarding and allowing for a potentially broad application.
        \item The dynamic generation of the challenges prevents users from preparing or sharing the correct answers in advance, reducing the risk of credential sharing.
    \end{itemize}
    \item[] \textit{Liabilities:}
    \begin{itemize}[topsep=0pt]
        \item The legitimate user may not recall the details needed to answer dynamically generated challenges, significantly limiting authentication accuracy and usability.
        \item The use of personal data from external sources raises significant privacy and data protection concerns and may be subject to legal or regulatory restrictions.
        \item If there is no internal historical or user data from the external data sources used, authentication with this pattern is not possible.
 
    \end{itemize}
\end{itemize}

\noindent \textbf{Known Uses:} In the U.S., credit bureaus such as Equifax~\cite{equifax} and Experian~\cite{experian}, as well as data aggregation and collection companies such as Veratad~\cite{veratad_kba} and LexisNexis~\cite{lexisnexis_instantid_qa} offer dynamic KBA solutions. Outside the U.S., adoption is significantly constrained by data protection regulations, such as those in the EU. \newpage

\pattern{Shared Secrets-based Authentication}

\noindent \textbf{Context:} A system is designed for use in an environment where subsidiary authentication is required for a primary authentication pattern, but where neither specialized hardware nor access to external personal data sources is available or practical. To meet this requirement, a solution is needed that enables the system to verify a user's identity using personal knowledge the user establishes during an upfront configuration step.

\noindent \textbf{Problem:} How can a system further ensure a user's identity without additional hardware, using personal knowledge that the user configures themselves upfront?  

\noindent \textbf{Forces:} 
\begin{itemize}[leftmargin=1em, topsep=0pt]
    \item[-] \textit{Third-party independence vs. Onboarding overhead}: User-configured secrets eliminate the need for external data sources or an issuance infrastructure, thereby minimizing privacy risks; however, the configuration step required for this makes the onboarding process more time-consuming, affecting usability and acceptance of the technique.
    \item[-] \textit{Low memorization burden vs. High vulnerability:} By allowing users to set their own secrets, they have the option to choose information that is easily memorizable; however, personal information is often publicly available or can be obtained through social engineering, which significantly reduces the associated level of security.
    \item[-] \textit{Infrastructure independence vs. Compromise persistence:} Preconfigured secrets do not require any infrastructure for dynamic generation and are easily verifiable; however, they are generally vulnerable to compromise, as they are purely static and remain a vulnerability after a successful compromise until the user explicitly changes them.
\end{itemize}

\noindent \textbf{Solution:} Use one or more personal knowledge challenges (shared secrets) configured by the user as an additional authenticator. During a dedicated configuration step, the user defines the shared secrets and provides the corresponding answers, which the system stores as a reference. For authentication, the system asks the user for one or more of the shared secrets and prompts them to provide the corresponding answers. If the answers provided match the stored reference, the user is authenticated and granted access~\cite{Rabadao.2023}.

To prevent unauthorized access, the system should apply a lockout policy after a limited number of failed authentication attempts and require a minimum number of correctly answered questions before granting access.

\noindent \textbf{Consequences:}

\begin{itemize}[leftmargin=0pt, topsep=0pt]
    \item[] \textit{Benefits:}
    \begin{itemize}[topsep=0pt]
        \item As the user configures the secrets, this pattern can provide subsidiary authentication without relying on external infrastructure.
        \item The user retains control over the shared secrets, alleviating data privacy concerns and making this pattern suitable in more strongly regulated environments.
    \end{itemize}
    \item[] \textit{Liabilities:}
    \begin{itemize}[topsep=0pt]
        \item Highly vulnerable to social engineering attacks, as preconfigured challenges can be guessed through mining publicly available information~\cite{Alomar.2017}.
        \item Depending on the challenges set offered, users may struggle to provide answers that they can easily recall later, limiting usability.
        \item Configured challenges and answers are static; once compromised, they remain a vulnerability until the user changes them.
        \item The configuration step adds user involvement to the onboarding process, potentially reducing user adoption.
    \end{itemize}
\end{itemize}

\noindent \textbf{Known Uses:} Widely used for account recovery and password reset across web applications~\cite{Reeder.2011}. \newpage

\section{Possession-based Authentication Patterns}\label{sec:possession}

With possession-based authentication patterns, subjects authenticate themselves by proving possession of and control over a physical token or device. As with knowledge-based patterns, this class is not limited to human subjects, and authentication data derived from possession-based patterns is less sensitive than that from inherence-based patterns, allowing transmission to remote systems. Furthermore, they do not require humans to memorize secrets. This makes them resilient to memory-related limitations and phishing attacks. Lastly, possession-based authentication patterns offer inherently high resistance to network-based interception and credential theft, since the credentials are stored on the possession-based token itself and are, thus, not exposed during transmission~\cite{Pahuja.2024}. 

However, possession-based authentication patterns also share common liabilities. Unlike inherence-based patterns, the authenticator can be separated from the subject’s identity. As a result, while they are inherently resistant to network-based threats, vulnerability in possession-based authentication patterns shifts the risk to physical-world threats, as physical authenticators can be easy targets for loss and theft \cite{Pahuja.2024}. This requires additional authentication patterns or security measures to protect against misuse. Furthermore, possession-based patterns result in higher costs due to organizational and infrastructural overhead. They require issuing or registering the token or device before authentication. This creates a dependency on external infrastructure for registration, whether a PKI, a card issuer, or a platform provisioning service, and adds operational overhead, which limits overall applicability.

\subsection{Primary Authentication} We identified two primary possession-based authentication patterns: smart card and passkey authentication. 

\addvspace{10pt}

\pattern{Smart Card Authentication}

\noindent \textbf{Context:} A system is designed for use in an environment with high security, privacy, and confidentiality requirements, and in which users must be able to authenticate themselves both locally and remotely across multiple devices or access points. To meet these requirements, a solution is needed that binds the authentication credential to a dedicated physical token that a central authority can issue, manage, and revoke.

\noindent \textbf{Problem:} How can a system enable users to authenticate themselves across multiple devices and access points using a dedicated physical token that is managed by a central authority?

\noindent \textbf{Forces:} 
\begin{itemize}[leftmargin=1em, topsep=0pt]
    \item[-] \textit{Authority control vs. Administrative overhead:} A central management and issuance authority enables systematic, centralized control over the lifecycle of authenticators and certificates, offering advantages for revoking lost or compromised credentials; however, this comes at the cost of significant administrative overhead for day-to-day operations.
    \item[-] \textit{Broad accessibility via access points vs. Hardware infrastructure burden:} Storing credentials on a physical device allows them to be used at all access points, possibly even worldwide, without requiring an additional enrollment; however, this requires the deployment and maintenance of compatible readers at each access point, which entails significant upfront infrastructure investments and a high degree of complexity during deployment.
    \item[-] \textit{Usability vs. Environmental degradation:} Scanning a physical token is generally a usable method for authentication for a user; however, this high level of usability can be significantly compromised over time due to reduced readability caused by environmental factors.
\end{itemize}

\noindent \textbf{Solution:} Use a dedicated physical token in the form of a smart card to store the authentication credential. During an enrollment process managed by a central authority, the authority issues a smart card to the user, generates the authentication credential, stores it on the card, and registers the corresponding certificate with the central authority. To authenticate, the user presents the smart card to a compatible reader. The reader reads the authentication credential from the card and transmits it to the system for verification. If the system successfully verifies the authentication credential against the registered certificate, it authenticates the user and grants access~\cite{Mayes.2007}.

To prevent unauthorized access, the system should require the user to provide an additional knowledge-based factor in high-security environments, for example, by using the \textsc{7. Authority-issued PIN Authentication} pattern to verify the smart card before the authentication credential can be read~\cite{Rankl.2007}. \pagebreak

\noindent \textbf{Consequences:}

\begin{itemize}[leftmargin=0pt, topsep=0pt]
    \item[] \textit{Benefits:}
    \begin{itemize}[topsep=0pt]
        \item Smart cards are highly resistant to credential theft, as the credential is stored on the card and is not exposed during transmission.
        \item The central authority’s full control over the authentication credentials allows their immediate revocation in the event of loss, theft, or misuse.
        \item Smart cards are compatible with various devices and access points without requiring re-registration. This makes them particularly well-suited for large-scale deployments across organizational boundaries.
        \item Smart cards can store additional data, such as biometric data, opening up a wide range of possible applications~\cite{Waldmann.2012}.
    \end{itemize}
    \item[] \textit{Liabilities:}
    \begin{itemize}[topsep=0pt]
        \item Smart cards require specialized reader hardware at each access point, which results in significant infrastructure costs and a complex deployment process.
        \item A lost or stolen smart card grants the finder physical access to its credential, making additional security measures inevitable.
        \item The process of issuing the smart card adds organizational overhead and is significantly more inconvenient for the user than alternative patterns.
        \item Smart cards are subject to physical wear and tear and damage, which can compromise authentication accuracy.
        \item The dependency on a central authority for issuance and certificate management introduces a single point of failure.
        \item When smart cards carry additional personal or biometric information, the pattern raises additional high privacy and security concerns and risks, requiring additional privacy-preserving controls~\cite{Waldmann.2012}.
    \end{itemize}
\end{itemize}

\noindent \textbf{Known Uses:} EMV cards in retail banking at point-of-sale terminals and ATMs are protected by an authority-issued PIN~\cite{Renaud.2015}. Electronic government~\cite{SmartCardAlliance, PIV} and identity cards~\cite{Waldmann.2012}. Physical access control in high-security environments, such as banking and healthcare facilities~\cite{Mayes.2007}. \newpage



\pattern{Passkey Authentication}

\noindent \textbf{Context:} A system is designed for use in an environment that requires large-scale authentication of remote users, where the effort required to manage an infrastructure with physical tokens is disproportionately high, and where knowledge-based authentication is considered inadequate due to its vulnerability to phishing and the theft of credentials. A solution is needed that enables strong, phishing-resistant authentication without distributing or managing dedicated physical tokens.

\noindent \textbf{Problem:} How can a system authenticate remote users in a phishing-resistant manner without relying on shared secrets and without requiring an infrastructure that uses dedicated physical tokens?

\noindent \textbf{Forces:} 
\begin{itemize}[leftmargin=1em, topsep=0pt]
    \item[-] \textit{Large-scale deployability vs. Ecosystem dependency:} Managing credentials without an external issuing authority or physical token infrastructure significantly reduces operational overhead and can therefore be deployed on a large scale; however, such an approach typically requires the synchronization of key material across a user’s entire platform ecosystem, which creates a dependency on that specific platform provider.
    \item[-] \textit{Phishing resistance vs. Device dependency:} Storing credentials locally as a key pair instead of a shared secret protects against phishing and the risk of interception; however, this ties authentication to possession of that specific device, meaning that if the device is lost or replaced, the user may no longer be able to authenticate without a separate recovery method.
    \item[-] \textit{Device-based security vs. User base coverage:} To achieve phishing resistance without relying on shared secrets, the user’s device itself must generate, store, and process the cryptographic key material; however, this requires device and platform capabilities that not all devices necessarily have, which may exclude users with unsupported devices and limit the target user base that the system can reach.
\end{itemize}

\noindent \textbf{Solution:} Use a platform-internal asymmetric key pair as proof of authentication. In a dedicated registration step, the user initiates the device to generate a key pair and transmit the public key to the system, which stores it as a reference for future authentication. The user's device stores the private key securely. To authenticate, the system sends a challenge to the user's device. The device signs the challenge with the private key and returns the signature to the system for verification. If the system successfully verifies the signature using the registered public key, it authenticates the user and grants access~\cite{George.2024}.

To prevent unauthorized access, the system should require the user to perform a local verification step, such as with the \textsc{1. Fingerprint Authentication} or \textsc{6. Self-managed PIN Authentication} pattern, before the device signs the challenge.

\noindent \textbf{Consequences:}

\begin{itemize}[leftmargin=0pt, topsep=0pt]
    \item[] \textit{Benefits:}
    \begin{itemize}[topsep=0pt]
        \item Passkeys are inherently resistant to phishing attacks, rendering them much more secure than knowledge-based authentication patterns.
        \item The system manages the entire lifecycle of authentication data independently, without relying on an external issuing authority or an infrastructure that uses physical tokens.
        \item Passkeys do not require users to memorize or manage secrets, reducing cognitive burden and eliminating the need for credential recovery mechanisms.
        \item Passkeys do not require distributing or managing physical tokens, reducing operational overhead and making large-scale deployment more practical.
    \end{itemize}
    \item[] \textit{Liabilities:}
    \begin{itemize}[topsep=0pt]
        \item Passkeys are bound to the user's device, so authentication credentials are lost if the device is lost or malfunctions. 
        \item While passkeys can be synced to reduce dependence on the user's device, this creates a dependency on the user's platform ecosystem (e.g., iCloud, Google Password Manager), which can conflict with enterprise security policies or regulatory requirements.
        \item Passkeys require the user's device and platform to support the underlying authentication standard (WebAuthn and FIDO2), limiting their applicability in environments with older or limited-capability devices.
    \end{itemize}
\end{itemize}

\noindent \textbf{Known Uses:} Common and emerging authentication pattern for web applications, such as for Google, Apple, and Microsoft account authentication, or authentication in GitHub~\cite{Bhardwaj.2026}. \newpage


\subsection{Subsidiary Authentication}
We identified three subsidiary possession-based authentication patterns: hardware-token one-time password (OTP), software-token OTP, and out-of-band OTP authentication.

\addvspace{10pt}

\pattern{Hardware-Token OTP Authentication}

\noindent \textbf{Context:} A system is designed for use in an environment with high security requirements, where subsidiary authentication for a primary authentication pattern is required. To meet these requirements, the technique used to generate the subsidiary authentication credentials must be tamper-resistant and operate independently of the user's general-purpose devices, ensuring it cannot be compromised by malware or unauthorized access to those devices.

\noindent \textbf{Problem:} How can a system provide subsidiary, phishing-resistant authentication using a tamper-resistant approach that operates independently of the user's general-purpose devices?

\noindent \textbf{Forces:} 
\begin{itemize}[leftmargin=1em, topsep=0pt]
    \item[-] \textit{Dedicated hardware device vs. Operational overhead:} The token's independent operability from the user's general-purpose devices ensures that its secret cannot be compromised by malware or unauthorized access to the device; however, this independence requires the token to be physically manufactured, issued to each user in advance, and later reissued or deactivated in the event of loss, theft, or damage, yielding additional logistical and operational costs and, thus, limiting scalability in large or geographically distributed deployments.
    \item[-] \textit{Broad compatibility vs. Carry burden:} By generating login credentials, the token can, in principle, be used with any authentication endpoint without requiring special readers or client-side software; however, this requires the user to have the appropriate device on hand and to actively operate it at the time of authentication, which they might forget to do and which could disrupt their workflow.
    \item[-] \textit{Lightweight verification vs. Provider-level exposure:} Such a technique requires that the login credentials be derived from a shared secret, which allows the system to verify it independently without requiring the additional architecture needed for certificate- or key-pair-based approaches; however, if this shared secret is compromised at the system or provider level, all login credentials derived from it are simultaneously compromised, resulting in a system-wide single point of failure.
\end{itemize}

\noindent \textbf{Solution:} Use a dedicated, physical hardware token that generates a dynamic, short-lived one-time password (OTP). Before issuing the token, the issuing authority registers the shared secret, pre-stored on the hardware token, in the system. The token is then issued to the user. To authenticate, the user uses the hardware token, which generates an OTP and transmits it to the system for verification. If the transmitted OTP matches the value that the system has independently derived from the shared secret, the user is authenticated and granted access~\cite{Lu.2008}.

To prevent unauthorized access, the system should enforce a strict validity window for each OTP and reject all login credentials that have already been used. 

\noindent \textbf{Consequences:} 

\begin{itemize}[leftmargin=0pt, topsep=0pt]
    \item[] \textit{Benefits:}
    \begin{itemize}[topsep=0pt]
        \item OTP credentials are dynamic and short-lived, so that any credentials that are intercepted or stolen become unusable after a single use or once their validity period expires.
        \item Hardware tokens are resistant to threats posed by credential theft via malware or device compromise.
        \item Hardware OTP tokens are compatible with any authentication endpoint that accepts alphanumeric input, so neither dedicated readers nor client software are required at the authentication endpoint.
    \end{itemize}
    \item[] \textit{Liabilities:}
    \begin{itemize}[topsep=0pt]
        \item Hardware tokens can be lost, stolen, or physically damaged, requiring the issuing authority to implement processes for reissuing and revoking them.
        \item Hardware tokens have limited usability, as users may forget to carry them and the process of generating OTPs can disrupt their workflow.
        \item Hardware tokens must be physically distributed to users in advance, involving costs and logistical challenges that limit scalability in large or geographically distributed deployments.
        \item The shared secret poses a long-term security vulnerability: if it is compromised at the system or vendor level, all tokens with this secret are affected.
    \end{itemize}
\end{itemize}

\noindent \textbf{Known Uses:} RSA SecurID tokens~\cite{Todorov.2007} are frequently used for authentication on corporate VPNs and for remote access. OATH- and FIDO2-compliant hardware tokens from vendors such as Yubico, Feitian, and SoloKeys~\cite{Casagrande.2025}. 

\pattern{Software-Token OTP Authentication}

\noindent \textbf{Context:} A system is designed for use in an environment with a broad user base and where subsidiary authentication is required for a primary authentication pattern. To meet these requirements, the technique for generating authentication credentials must be deployable without pre-distributing dedicated physical authenticators to users.

\noindent \textbf{Problem:} How can a system provide subsidiary authentication that is applicable to a broad user base by using the user's personal device to generate authentication credentials?

\noindent \textbf{Forces:} 
\begin{itemize}[leftmargin=1em, topsep=0pt]
    \item[-] \textit{Broad deployability vs. Device integrity:} Since the requirements are limited to a personal device and an Internet connection, the system can be used by a large, diverse user base without the need for prior distribution; however, the system offers no guarantee of the integrity of the devices that generate the login credentials, and this uncertainty increases as the user base grows and becomes more diverse.
    \item[-] \textit{Low operational costs vs. Broad attack surface:} Using a personal device of the user eliminates the effort of issuing, replacing, and retrieving hardware, reducing operating costs; however, that device usually runs many other applications that are outside of the system's control, exposing the shared secret to a broader range of threats.
    \item[-] \textit{High usability vs. Compliance:} Allowing users to self-enroll at any time from any device and any location ensures a quick and seamless registration process; however, this process does not provide the system with verified assurance that the registration took place on a genuine, uncompromised device, which may conflict with regulatory or internal corporate compliance requirements.
\end{itemize}

\noindent \textbf{Solution:} Use a software-based authentication application installed on the user’s personal device to generate dynamic, short-lived one-time passwords (OTPs). During a dedicated registration step, the user installs the authentication application and uses it to exchange a shared secret with the system, which the system stores as a reference for future authentication. To authenticate, the user opens the authentication application, which generates an OTP from the shared secret and transmits it to the system for verification. If the transmitted OTP matches the value the system independently derives from the shared secret, the user is authenticated and granted access~\cite{Sun.2015}.

To prevent unauthorized access, the system should enforce a strict validity window for each OTP and reject any authentication credentials that have already been used.

\noindent \textbf{Consequences:}

\begin{itemize}[leftmargin=0pt, topsep=0pt]
    \item[] \textit{Benefits:}
    \begin{itemize}[topsep=0pt]
        \item Deployable to a broad user base, as registration requires only an internet connection and a personal device.
        \item Using a personal device avoids the logistical effort involved in issuing, replacing, and revoking physical tokens, significantly reducing operational costs.
        \item Because the user uses a personal device, they can set up the software at any time and from any location, making this subsidiary pattern quick and easy to use.
    \end{itemize}
    \item[] \textit{Liabilities:}
    \begin{itemize}[topsep=0pt]
        \item Malware, device compromise, or unauthorized access can extract the shared secret.
        \item Unlike dedicated hardware tokens, a general-purpose device has a broader attack surface as it runs multiple applications and is exposed to a wider range of threats.
        \item This pattern does not provide the relying party with any assurance regarding the integrity of the devices on which the access credentials are generated---a risk that increases with the size and heterogeneity of the user base.
    \end{itemize}
\end{itemize}

\noindent \textbf{Known Uses:} Software-token OTP implementations can be found in a wide range of applications, such as Google Authenticator, Microsoft Authenticator,  Twilio Authy Authenticator, and Sophos Authenticator~\cite{Ang.2025}. \newpage

\pattern{Out-of-Band OTP Authentication}

\noindent \textbf{Context:} A system is designed for use in an environment with a broad, potentially unregistered user base, where subsidiary authentication for a primary authentication pattern is required, and where at least one out-of-band communication channel to any user can be assumed. To meet this requirement, a solution is needed that provides subsidiary authentication by employing the out-of-band channel.

\noindent \textbf{Problem:} How can a system provide subsidiary authentication to a broad, potentially unregistered user base by using an out-of-band communication channel the user already has access to?

\noindent \textbf{Forces:} 
\begin{itemize}[leftmargin=1em, topsep=0pt]
    \item[-] \textit{Broad user base coverage vs. Third-party dependency:} By reusing a channel that the user already has, it is possible to reach a broad user base, including non-registered users, without requiring a special registration infrastructure; however, the provision and security of the channel are completely outside of the system’s control, so its trustworthiness depends on third-party guarantees.
    \item[-] \textit{Low integration overhead vs. Channel-inherent vulnerability:} Reusing a channel that the user already has allows avoiding the introduction of new, dedicated credentials or infrastructure; however, the channel has its own pre-existing vulnerabilities, such as interception at the network level or compromise of the account that protects it, which the pattern inherits without being able to mitigate them directly.
    \item[-] \textit{User familiarity vs. Delivery reliability:} Since users are usually familiar with these out-of-band channels for other purposes as well, the authentication step is easy to understand; however, delivery cannot be guaranteed, as factors beyond the system’s control, such as network congestion, spam filters, or connection issues, can delay or prevent the delivery of the OTP, significantly impairing usability.
\end{itemize}

\noindent \textbf{Solution:} Let the user configure one or more out-of-band communication channels, such as a phone number or email address, in a dedicated registration step. The system stores them as reference destinations. To authenticate, the system generates a dynamic, short-lived one-time password (OTP), stores it (or a derived value) for verification, and transmits it to one of the registered destinations. The user retrieves the OTP from their out-of-band channel and submits it to the system. If the submitted OTP matches the stored value and has not expired, the user is authenticated and granted access~\cite{Botwright.2023, Oorschot.2020}.

To prevent unauthorized access, the system should enforce a strict validity window for each OTP, reject OTPs that have already been used, apply a lockout policy after a limited number of failed attempts, and monitor for anomalous request patterns that may indicate an attempt to intercept or flood a channel.

\noindent \textbf{Consequences:}

\begin{itemize}[leftmargin=0pt, topsep=0pt]
    \item[] \textit{Benefits:}
    \begin{itemize}[topsep=0pt]
        \item Deployable to broad user bases, as in many use cases, users can be expected to have access to at least an email address or a personal phone.
        \item Low registration efforts, especially in most web applications, where an email address must be provided to register anyway.
        \item Well understood by users due to widespread use.
    \end{itemize}
    \item[] \textit{Liabilities:}
    \begin{itemize}[topsep=0pt]
        \item The delivery channel is outside the system's control, making the pattern dependent on the security and availability of third-party or email providers.
        \item SMS delivery is vulnerable to SIM-swapping and SS7 interception attacks~\cite{Lei.2021}, undermining the trustworthiness of the out-of-band channel.
        \item The security of email transmission depends entirely on the security of the user's email account; if that account is compromised, the OTP channel is also at risk.
        \item Limited usability, as delivery of the OTP may be delayed or fail entirely due to network congestion, spam filters, or connection issues.
    \end{itemize}
\end{itemize}

\noindent \textbf{Known Uses:} Widely used as subsidiary authentication across banking, e-commerce, and social media platforms~\cite{Hsieh.2011, Aravindhan.2013}, especially for account verification and password reset. \newpage

\section{Conclusions \& Future Work}\label{sec:conclusion}

This paper presents a catalog of 14 distinct user authentication patterns. The catalog is organized by two properties: the pattern's authentication factor and its usual role in the authentication context. With these findings, we aim to make a significant contribution toward improving the practical application of security patterns.

Along with the detailed pattern descriptions, we also sketched authentication factor-level properties that are inherent to the authentication patterns of a given factor. However, this also reveals a limitation of this paper. It indicates that authentication factors exhibit pattern-like characteristics as well. Given the relationship between authentication factors and authentication patterns, properties inherent to authentication factors could be formulated in terms of abstract patterns, with the 14 patterns presented in our catalog serving as their concrete patterns.

A second limitation is that the catalog does not account for possible combinations of authentication patterns. While the entries in this catalog clearly represent authentication patterns, they represent single employments. In practice, however, using multiple authentication patterns together to form a standalone authentication technique is common. For example, many web applications implement two-factor authentication using \textsc{5. Text Password Authentication} for primary authentication with \textsc{13. Software-token OTP Authentication} for subsidiary authentication.

We plan to address both limitations---the factor-pattern level and pattern-based multiple employments of authentication patterns---in future work. To this end, we plan to extend the authentication patterns by a suitable authentication pattern language. We also plan to extend the catalog of user authentication patterns with patterns that can be identified for machine-to-machine authentication.


\bibliographystyle{unsrt}  
\bibliography{Bibliography}

@misc{OWASP.Top10,
  author       = {{The OWASP Foundation}},
  title        = {{OWASP Top Ten}},
  organization = {{Open Worldwide Application Security Project (OWASP)}},
  url = {{{https://owasp.org/Top10/2025/}}},
  year         = {2026},
  note         = {last accessed: 2026-05-08}
}

@inproceedings{Gressl.2019,
 author = {Gressl, Lukas and Steger, Christian and Neffe, Ulrich},
 title = {Consideration of Security Attacks in the Design Space Exploration of Embedded Systems},
 pages = {530--537},
 booktitle = {2019 22nd Euromicro Conference on Digital System Design (DSD)},
 year = {2019},
 doi = {10.1109/DSD.2019.00082}
}

@article{Furnell.2020,
  title={Addressing cyber security skills: the spectrum, not the silo},
  author={Furnell, Steven and Bishop, Matt},
  journal={Computer fraud \& security},
  volume={2020},
  number={2},
  pages={6--11},
  year={2020},
  publisher={MA Business London}
}

@inproceedings{Gutfleisch.2022,
 author = {Gutfleisch, Marco and Klemmer, Jan H. and Busch, Niklas and Acar, Yasemin and Sasse, M. Angela and Fahl, Sascha},
 title = {How Does Usable Security (Not) End Up in Software Products? Results From a Qualitative Interview Study},
 pages = {893--910},
 booktitle = {2022 IEEE Symposium on Security and Privacy (SP)},
 year = {2022},
 doi = {10.1109/SP46214.2022.9833756}
}

@inproceedings{Naji.2025,
 author = {Naji, Houda and Reichmann, Felix and Bruns, Tobias and Sasse, M. Angela and Naiakshina, Alena},
 title = {{``It's not my responsibility to write them'': an empirical study of software product managers and security requirements}},
 publisher = {{USENIX Association}},
 isbn = {978-1-939133-52-6},
 series = {SEC '25},
 booktitle = {Proceedings of the 34th USENIX Conference on Security Symposium},
 year = {2025},
 address = {USA}
}

@book{Schumacher.2006,
 author = {Schumacher, Markus and Fernandez-Buglioni, Eduardo and Hybertson, Duane and Buschmann, Frank and Sommerlad, Peter},
 year = {2006},
 title = {Security patterns : integrating security and systems engineering / Markus Schumacher, Eduardo Fernandez-Buglioni, Duane Hybertson, Frank Buschmann, Peter Sommerlad},
 address = {Chichester, England},
 publisher = {{John Wiley {\&} Sons}},
 isbn = {9780470858851}
}

@book{Fernandez-Buglioni.2013,
 author = {Fernandez-Buglioni, Eduardo},
 year = {2013},
 title = {Security Patterns in Practice: Designing Secure Architectures Using Software Patterns},
 address = {Chichester},
 publisher = {{Wiley {\&} Sons}},
 isbn = {9781119998945},
 series = {Wiley Software Patterns Series}
}

@inproceedings{Yskout.2015,
 author = {Yskout, Koen and Scandariato, Riccardo and Joosen, Wouter},
 title = {Do Security Patterns Really Help Designers?},
 pages = {292--302},
 volume = {1},
 booktitle = {2015 IEEE/ACM 37th IEEE International Conference on Software Engineering},
 year = {2015},
 doi = {10.1109/ICSE.2015.49}
}

@inproceedings{van.den.Berghe.2018,
 author = {{van den Berghe}, Alexander and Yskout, Koen and Joosen, Wouter},
 title = {Security patterns 2.0: Towards Security Patterns Based on Security Building Blocks},
 pages = {45--48},
 publisher = {{Association for Computing Machinery}},
 isbn = {9781450357272},
 series = {SEAD '18},
 booktitle = {Proceedings of the 1st International Workshop on Security Awareness from Design to Deployment},
 year = {2018},
 address = {New York, NY, USA},
 doi = {10.1145/3194707.3194715}
}

@inproceedings{van.den.Berghe.2022,
 author = {{van den Berghe}, Alexander and Yskout, Koen and Joosen, Wouter},
 title = {A reimagined catalogue of software security patterns},
 pages = {25--32},
 publisher = {{Association for Computing Machinery}},
 isbn = {9781450392907},
 series = {EnCyCriS '22},
 booktitle = {Proceedings of the 3rd International Workshop on Engineering and Cybersecurity of Critical Systems},
 year = {2022},
 address = {New York, NY, USA},
 doi = {10.1145/3524489.3527301}
}

@misc{Distrinet,
  author       = {{DistriNet}},
  title        = {{Security Pattern Catalog}},
  organization = {{imec-DistriNet research group, KU Leuven}},
  url = {{{https://securitypatterns.distrinet-research.be/}}},
  year         = {2026},
  note         = {last accessed: 2026-05-08}
}

@inproceedings{Yoder.1997,
 author = {{Joseph W. Yoder} and {Jeffrey Barcalow}},
 title = {Architectural Patterns for Enabling Application Security},
 url = {https://www.plopcon.org/pastplops/plop97/Proceedings/yoder.pdf},
 booktitle = {4th Pattern Languages of Programming Conference},
 year = {1997}
}

@article{Fernandez.2022,
 author = {Fernandez, Eduardo B. and Yoshioka, Nobukazu and Washizaki, Hironori and Yoder, Joseph},
 year = {2022},
 title = {Abstract security patterns and the design of secure systems},
 pages = {7},
 volume = {5},
 number = {1},
 issn = {2523-3246},
 journal = {Cybersecurity},
 doi = {10.1186/s42400-022-00109-w}
}

@techreport{TheOpenGroup,
  author       = {B. Blakeley and C. Heath and Members of the Open Group Security Forum},
  title        = {Technical Guide: Security Design Patterns},
  institution  = {The Open Group},
  year         = {2004},
  url          = {http://www.opengroup.org/bookstore/catalog/g031.htm},
  note         = {Accessed: 14 October 2025}
}

@inproceedings{Cordeiro.2022,
 author = {Cordeiro, Andr{\'e} and Vasconcelos, Andr{\'e} and Correia, Miguel},
 title = {A Catalog of Security Patterns},
 publisher = {{The Hillside Group}},
 isbn = {9781941652183},
 series = {PLoP '22},
 booktitle = {Proceedings of the 29th Conference on Pattern Languages of Programs},
 year = {2023},
 address = {USA}
}

@book{Todorov.2007,
  title={Mechanics of User Identification and Authentication: Fundamentals of Identity Management},
  author={Todorov, D.},
  isbn={9781040169469},
  year={2007},
  publisher={CRC Press}
}

@inproceedings{Heyman.2007,
 author = {Heyman, Thomas and Yskout, Koen and Scandariato, Riccardo and Joosen, Wouter},
 title = {An Analysis of the Security Patterns Landscape},
 pages = {3},
 booktitle = {Third International Workshop on Software Engineering for Secure Systems (SESS'07: ICSE Workshops 2007)},
 year = {2007},
 doi = {10.1109/SESS.2007.4}
}

@inproceedings{Catalog.Paper,
author={Alex R. Mattukat and Vincent Schmandt and Timo Langstrof and Michael Zerbe and Horst Lichter},
title={A Faceted Classification of Authenticator-Centric Authentication Techniques},
booktitle={Proceedings of the 21st International Conference on Evaluation of Novel Approaches to Software Engineering - Volume 1: ENASE},
year={2026},
pages={568-578},
publisher={SciTePress},
organization={INSTICC},
doi={10.5220/0014982200004015},
isbn={978-989-758-828-0},
issn={2184-4895},
}

@techreport{Yskout.2007,
    author = {Yskout, Koen and Heyman, Thomas and Scandariato, Riccardo and Joosen, Wouter},
    title = {A system of security patterns},
    institution = {DistriNet Research Group, Katholieke Universiteit Leuven},
    year = 2007,
    address = {Belgium}
}

@misc{NIST.800-63-4,
 author = {{David Temoshok} and {Diana Proud-Madruga} and {Yee-Yin Choong} and {Ryan Galluzzo} and {Sarbari Gupta} and {Connie LaSalle} and {Naomi Lefkovitz} and {Andrew Regenscheid}},
 date = {2025},
 title = {Digital Identity Guidelines: NIST Special Publication},
 number = {SP 800-63-4},
 institution = {{National Institute of Standards and Technology (NIST)}},
 doi = {10.6028/NIST.SP.800-63-4},
}

@book{Maltoni.2022,
 author = {Maltoni, Davide and Maio, Dario and Jain, Anil K. and Feng, Jianjiang},
 year = {2022},
 title = {Handbook of Fingerprint Recognition},
 edition = {3},
 publisher = {{Springer Nature}},
 isbn = {978-3-030-83624-5},
}

@unpublished{SMS,
  author    = {Mattukat, Alex R. and Langstrof, Timo and Zerbe, Michael and Lichter, Horst},
  title     = {What Are Authentication Patterns? A Systematic Mapping Study},
  note      = {Paper in preparation},
  year      = {2026},
}

@article{Langenderfer.2005,
author = {Langenderfer, Jeff and Linnhoff, Stefan},
title = {The Emergence of Biometrics and Its Effect on Consumers},
journal = {Journal of Consumer Affairs},
volume = {39},
number = {2},
pages = {314-338},
doi = {https://doi.org/10.1111/j.1745-6606.2005.00017.x},
url = {https://onlinelibrary.wiley.com/doi/abs/10.1111/j.1745-6606.2005.00017.x},
eprint = {https://onlinelibrary.wiley.com/doi/pdf/10.1111/j.1745-6606.2005.00017.x},
year = {2005}
}

@inproceedings{Alwahaishi.2020,
    author = { Alwahaishi, Saleh and Zdralek, Jaroslav },
    booktitle = { 2020 IEEE International Conference on Cloud Computing in Emerging Markets (CCEM) },
    title = {{ Biometric Authentication Security: An Overview }},
    year = {2020},
    volume = {},
    ISSN = {},
    pages = {87-91},
    doi = {10.1109/CCEM50674.2020.00027},
    url = {https://doi.ieeecomputersociety.org/10.1109/CCEM50674.2020.00027},
    publisher = {IEEE Computer Society},
    address = {Los Alamitos, CA, USA},
    month =Nov
}

@book{Vielhauer.2010,
 author = {Vielhauer, Claus},
 year = {2010},
 title = {Biometric User Authentication for IT Security: From Fundamentals to Handwriting},
 address = {New York, NY, USA},
 edition = {1},
 publisher = {{Springer New York}},
 isbn = {978-1-4419-3873-2},
 series = {Advances in Information Security},
 doi = {10.1007/0-387-28094-4}
}

@article{Kim.2021,
 author = {Kim, Ejin and Choi, Hyoung-Kee},
 year = {2021},
 title = {Security Analysis and Bypass User Authentication Bound to Device of Windows Hello in the Wild},
 pages = {6245306},
 volume = {2021},
 number = {1},
 journal = {Security and Communication Networks},
 doi = {10.1155/2021/6245306}
}

@inproceedings{Chowdhury.2017,
 author = {{Chowdhury, Mozammel and Gao, Junbin and Islam, Rafiqul}},
 title = {Biometric Authentication Using Facial Recognition},
 pages = {287--295},
 publisher = {{Springer International Publishing}},
 isbn = {978-3-319-59608-2},
 editor = {{Deng, Robert and Weng, Jian and Ren, Kui and Yegneswaran, Vinod}},
 booktitle = {Security and Privacy in Communication Networks},
 year = {2017},
 address = {Cham}
}

@article{Chakraborty.2014,
       author = {{Chakraborty}, Saptarshi and {Das}, Dhrubajyoti},
        title = "{An Overview of Face Liveness Detection}",
      journal = {arXiv e-prints},
         year = 2014,
        month = may,
          eid = {arXiv:1405.2227},
        pages = {arXiv:1405.2227},
          doi = {10.48550/arXiv.1405.2227},
archivePrefix = {arXiv},
       eprint = {1405.2227},
 primaryClass = {cs.CV},
       adsurl = {https://ui.adsabs.harvard.edu/abs/2014arXiv1405.2227C}
}

@ARTICLE{Gonzalez-Soler.2021,
  author={González-Soler, Lázaro Janier and Gomez-Barrero, Marta and Chang, Leonardo and Pérez-Suárez, Airel and Busch, Christoph},
  journal={IEEE Access}, 
  title={Fingerprint Presentation Attack Detection Based on Local Features Encoding for Unknown Attacks}, 
  year={2021},
  volume={9},
  number={},
  pages={5806-5820},
  doi={10.1109/ACCESS.2020.3048756}}

@book{Bowyer.2016,
 author = {Bowyer, K. W. and Burge, M. J.},
 year = {2016},
 title = {Handbook of Iris Recognition},
 edition = {2},
 publisher = {{Springer-Verlag London}},
 isbn = {9781447167846},
 series = {Advances in Computer Vision and Pattern Recognition}
}

@book{Li.2005,
 author = {Li, S. Z. and Jain, A. K.},
 year = {2005},
 title = {Handbook of Face Recognition},
 publisher = {Springer},
 isbn = {9780387405957}
}

@book{Fitzgerald.2018,
 author = {Fitzgerald, T.},
 year = {2018},
 title = {CISO COMPASS: Navigating Cybersecurity Leadership Challenges with Insights from Pioneers},
 publisher = {{CRC Press}},
 isbn = {9780429677830}
}

@misc{TypingDNA,
  author       = {{TypingDNA}},
  title        = {{ActiveLock: Continuous Endpoint Authentication Using Typing Biometrics}},
  howpublished = {\url{https://www.typingdna.com/activelock-continuous-authentication}},
  note         = {Accessed: 2026-06-30},
  organization = {TypingDNA}
}

@misc{biocatch,
  author       = {BioCatch},
  title        = {{BioCatch} - {Behavioral} {Biometrics} to {Prevent} {Fraud} \& {Build} {Trust}},
  howpublished = {\url{https://www.biocatch.com/}},
  note         = {Accessed: 2026-06-30},
  organization = {BioCatch},
  year         = {2026}
}

@misc{behaviosec,
  author       = {{LexisNexis Risk Solutions}},
  title        = {{BehavioSec}® — A {Real-Time} {Behavioral} and {Device} {Intelligence} {Solution}},
  howpublished = {\url{https://risk.lexisnexis.com/global/en/products/behaviosec}},
  note         = {Accessed: 2026-06-30},
  organization = {LexisNexis Risk Solutions},
  year         = {2026}
}

@misc{meazurelearning,
  author       = {{Meazure Learning}},
  title        = {{Meazure Learning} - {A} {Full-Service} {Test} {Development}, {Delivery}, and {Proctoring} {Provider}},
  howpublished = {\url{https://www.meazurelearning.com/}},
  note         = {Accessed: 2026-06-30},
  organization = {Meazure Learning},
  year         = {2026}
}

@misc{pearsonvue,
  author       = {{Pearson Professional Assessments}},
  title        = {{Certification} {Exams} \& {Licensure} {Testing} | {Pearson} {Professional} {Testing}},
  howpublished = {\url{https://www.pearsonvue.com/us/en/test-takers.html}},
  note         = {Accessed: 2026-06-30. Formerly known as Pearson VUE},
  organization = {Pearson},
  year         = {2026}
}

@misc{aetna,
  author       = {Lee, Justin},
  title        = {Aetna Rolls Out {FIDO}, Behavioral Authentication for Healthcare Data Security},
  howpublished = {\url{https://www.biometricupdate.com/201707/aetna-rolls-out-fido-behavioral-authentication-for-healthcare-data-security}},
  note         = {Accessed: 2026-06-30},
  organization = {Biometric Update},
  month        = jul,
  year         = {2017}
}

@inproceedings{Bertok.2016,
  author={Bertók, Kornél and Fazekas, Attila},
  booktitle={2016 7th IEEE International Conference on Cognitive Infocommunications (CogInfoCom)}, 
  title={Face recognition on mobile platforms}, 
  year={2016},
  volume={},
  number={},
  pages={000037-000042},
  doi={10.1109/CogInfoCom.2016.7804521}
  }

@misc{alcatrazai,
  author       = {{Alcatraz AI}},
  title        = {{AI} Facial Authentication Access Control System | {Alcatraz} {AI}},
  howpublished = {\url{https://www.alcatraz.ai/}},
  note         = {Accessed: 2026-06-30},
  organization = {Alcatraz AI},
  year         = {2026}
}

@inproceedings{Bhagavatula.2015,
  title={Biometric authentication on iphone and android: Usability, perceptions, and influences on adoption},
  author={Bhagavatula, Rasekhar and Ur, Blase and Iacovino, Kevin and Kywe, Su Mon and Cranor, Lorrie Faith and Savvides, Marios},
  booktitle={USEC ’15: Workshop on Usable Security},
  year={2015},
  publisher={Internet Society},
  pages={1-10}
}

@inproceedings{Sharma.2018,
  author={Sharma, Lokesh and Mathuria, Manish},
  booktitle={2018 2nd International Conference on Inventive Systems and Control (ICISC)}, 
  title={Mobile banking transaction using fingerprint authentication}, 
  year={2018},
  volume={},
  number={},
  pages={1300-1305},
  doi={10.1109/ICISC.2018.8399016}}

@article{Jo.2016,
author = {Jo, Young-Hoo and Jeon, Seong-Yun and Im, Jong-Hyuk and Lee, Mun-Kyu},
title = {Security Analysis and Improvement of Fingerprint Authentication for Smartphones},
journal = {Mobile Information Systems},
volume = {2016},
number = {1},
pages = {8973828},
doi = {https://doi.org/10.1155/2016/8973828},
url = {https://onlinelibrary.wiley.com/doi/abs/10.1155/2016/8973828},
eprint = {https://onlinelibrary.wiley.com/doi/pdf/10.1155/2016/8973828},
year = {2016}
}

@article{Labati.2017,
author = {Labati, Ruggero Donida and Genovese, Angelo and Mu\~{n}oz, Enrique and Piuri, Vincenzo and Scotti, Fabio and Sforza, Gianluca},
title = {Biometric Recognition in Automated Border Control: A Survey},
year = {2016},
issue_date = {June 2017},
publisher = {Association for Computing Machinery},
address = {New York, NY, USA},
volume = {49},
number = {2},
issn = {0360-0300},
url = {https://doi.org/10.1145/2933241},
doi = {10.1145/2933241},
journal = {ACM Comput. Surv.},
month = jun,
articleno = {24},
numpages = {39}
}

@inproceedings{Taleb.2014,
  author={Taleb, Imene and Amine Ouis, Mohamed El and Mammar, Madani Ould},
  booktitle={2014 4th International Symposium ISKO-Maghreb: Concepts and Tools for knowledge Management (ISKO-Maghreb)}, 
  title={Access control using automated face recognition: Based on the PCA \& LDA algorithms}, 
  year={2014},
  volume={},
  number={},
  pages={1-5},
  doi={10.1109/ISKO-Maghreb.2014.7033455}}

@book{Rathgeb.2012,
 author = {Rathgeb, Christian and Uhl, Andreas and Wild, Peter},
 year = {2012},
 title = {Iris Biometrics: From Segmentation to Template Security},
 publisher = {{Springer New York}},
 isbn = {9781461455714},
 series = {Advances in Information Security}
}

@article{Murad.2021,
 author = {{Mohammed Murad}},
 year = {2021},
 title = {Key markets for iris biometrics},
 url = {https://www.sciencedirect.com/science/article/pii/S0969476521000825},
 pages = {5--7},
 volume = {2021},
 number = {7},
 issn = {0969-4765},
 journal = {Biometric Technology Today},
 doi = {10.1016/S0969-4765(21)00082-5}
}

@article{Wang.2025,
author = {Wang, Mengdi and Bozkir, Efe and Kasneci, Enkelejda},
title = {Iris Style Transfer: Enhancing Iris Recognition with Style Features and Privacy Preservation through Neural Style Transfer},
year = {2025},
issue_date = {June 2025},
publisher = {Association for Computing Machinery},
address = {New York, NY, USA},
volume = {8},
number = {2},
url = {https://doi.org/10.1145/3729413},
doi = {10.1145/3729413},
journal = {Proc. ACM Comput. Graph. Interact. Tech.},
month = may,
articleno = {21},
numpages = {21}
}

@inproceedings{Conklin.2004,
  author={Conklin, A. and Dietrich, G. and Walz, D.},
  booktitle={37th Annual Hawaii International Conference on System Sciences, 2004. Proceedings of the}, 
  title={Password-based authentication: a system perspective}, 
  year={2004},
  volume={},
  number={},
  pages={10 pp.-},
  doi={10.1109/HICSS.2004.1265412}}

@inproceedings{Zezschwitz.2013,
author = {von Zezschwitz, Emanuel and Dunphy, Paul and De Luca, Alexander},
title = {Patterns in the wild: a field study of the usability of pattern and pin-based authentication on mobile devices},
year = {2013},
isbn = {9781450322737},
publisher = {Association for Computing Machinery},
address = {New York, NY, USA},
url = {https://doi.org/10.1145/2493190.2493231},
doi = {10.1145/2493190.2493231},
booktitle = {Proceedings of the 15th International Conference on Human-Computer Interaction with Mobile Devices and Services},
pages = {261–270},
numpages = {10},
location = {Munich, Germany},
series = {MobileHCI '13}
}

@article{Kruzikova.2022,
 author = {{Agata Kruzikova} and {Lenka Knapova} and {David Smahel} and {Lenka Dedkova} and {Vashek Matyas}},
 year = {2022},
 title = {Usable and secure? User perception of four authentication methods for mobile banking},
 url = {https://www.sciencedirect.com/science/article/pii/S0167404822000025},
 pages = {102603},
 volume = {115},
 issn = {0167-4048},
 journal = {Computers {\&} Security},
 doi = {10.1016/j.cose.2022.102603}
}

@inproceedings{Renaud.2015,
  author={Renaud, Karen and Volkamer, Melanie},
  booktitle={2015 World Congress on Internet Security (WorldCIS)}, 
  title={Exploring mental models underlying PIN management strategies}, 
  year={2015},
  volume={},
  number={},
  pages={18-23},
  doi={10.1109/WorldCIS.2015.7359406}}

@techreport{SmartCardAlliance,
  author      = {{Smart Card Alliance}},
  title       = {Using Smart Cards for Secure Physical Access},
  institution = {Smart Card Alliance},
  year        = {2003},
  month       = jul,
  number      = {ID-03003},
  address     = {Princeton Junction, NJ},
  url         = {https://www.securetechalliance.org/resources/pdf/Physical_Access_Report.pdf},
  note        = {A Smart Card Alliance Report}
}

@techreport{PIV,
  author      = {Dray, Jim and Corcoran, David},
  title       = {Personal Identity Verification Card Management Report},
  institution = {National Institute of Standards and Technology},
  type        = {NIST Interagency Report (NISTIR)},
  number      = {7284},
  address     = {Gaithersburg, MD},
  year        = {2006},
  month       = jan,
  url         = {http://piv.nist.gov}
}

@inproceedings{Chakraborty.2019,
author = {Chakraborty, Dhiman and Hanzlik, Lucjan and Bugiel, Sven},
title = {{simTPM}: User-centric {TPM} for Mobile Devices},
booktitle = {28th USENIX Security Symposium (USENIX Security 19)},
year = {2019},
isbn = {978-1-939133-06-9},
address = {Santa Clara, CA},
pages = {533--550},
url = {https://www.usenix.org/conference/usenixsecurity19/presentation/chakraborty},
publisher = {USENIX Association},
month = aug
}

@misc{equifax,
  author       = {{Equifax Inc.}},
  title        = {Digital Identity Trust},
  howpublished = {\url{https://www.equifax.com/business/product/digital-identity-trust/}},
  note         = {Accessed: 2026-06-30},
  organization = {Equifax}
}

@misc{experian,
  author       = {Funicelli, Brian},
  title        = {A Guide to User Authentication Types and Methods},
  howpublished = {\url{https://www.experian.com/blogs/insights/guide-user-authentication-types-methods/}},
  year         = {2024},
  month        = dec,
  note         = {Experian Insights blog; accessed 2026-06-30},
  organization = {Experian}
}

@misc{veratad_kba,
  author       = {{Veratad Technologies, LLC}},
  title        = {Knowledge-Based Authentication},
  howpublished = {\url{https://veratad.com/methods/knowledge-based-authentication}},
  note         = {Accessed: 2026-06-30},
  organization = {Veratad}
}

@misc{lexisnexis_instantid_qa,
  author       = {{LexisNexis Risk Solutions}},
  title        = {{InstantID}® {Q\&A}: Secure Knowledge Based Authentication},
  howpublished = {\url{https://risk.lexisnexis.com/products/instantid-q-and-a}},
  note         = {Accessed: 2026-06-30},
  organization = {LexisNexis Risk Solutions}
}

@inproceedings{Lei.2021,
  title={On the insecurity of SMS one-time password messages against local attackers in modern mobile devices},
  author={Lei, Zeyu and Nan, Yuhong and Fratantonio, Yanick and Bianchi, Antonio},
  booktitle={Network and Distributed Systems Security (NDSS) Symposium 2021},
  year={2021}
}

@ARTICLE{Alomar.2017,
  author={Alomar, Noura and Alsaleh, Mansour and Alarifi, Abdulrahman},
  journal={IEEE Communications Surveys \& Tutorials}, 
  title={Social Authentication Applications, Attacks, Defense Strategies and Future Research Directions: A Systematic Review}, 
  year={2017},
  volume={19},
  number={2},
  pages={1080-1111},
  doi={10.1109/COMST.2017.2651741}}

@book{Rabadao.2023,
 author = {Rabad{\~a}o, C. and Santos, L. and Costa, R.L.C.},
 year = {2023},
 title = {Information Security and Privacy in Smart Devices: Tools, Methods, and Applications: Tools, Methods, and Applications},
 publisher = {{IGI Global}},
 isbn = {9781668459935},
 series = {Advances in Information Security, Privacy, and Ethics}
}

@article{Reeder.2011,
  author={Reeder, Robert and Schechter, Stuart},
  journal={IEEE Security \& Privacy}, 
  title={When the Password Doesn't Work: Secondary Authentication for Websites}, 
  year={2011},
  volume={9},
  number={2},
  pages={43-49},
  doi={10.1109/MSP.2011.1}}

@book{Mayes.2007,
 author = {Mayes, K. and Markantonakis, K.},
 year = {2008},
 title = {Smart Cards, Tokens, Security and Applications},
 publisher = {{Springer US}},
 isbn = {9780387721989},
 series = {Computer Science}
}

@article{Waldmann.2012,
author={Waldmann, Ulrich and Türpe, Sven and Poller, Andreas and Vowé, Sven},
journal={ IEEE Security \& Privacy },
title={{ Electronic Identity Cards for User Authentication—Promise and Practice }},
year={2012},
volume={10},
number={01},
ISSN={1558-4046},
pages={46-54},
doi={10.1109/MSP.2011.148},
url = {https://doi.ieeecomputersociety.org/10.1109/MSP.2011.148},
publisher={IEEE Computer Society},
address={Los Alamitos, CA, USA},
month=jan}

@article{George.2024,
 author = {George, A. Shaji},
 year = {2024},
 title = {The Dawn of Passkeys: Evaluating a Passwordless Future},
 pages = {202--220},
 volume = {2},
 number = {1},
 journal = {Partners Universal Innovative Research Publication},
 doi = {10.5281/zenodo.10697886}
}

@inproceedings{Bhardwaj.2026,
 author = {Bhardwaj, Prince and Sastry, Nishanth},
 title = {State of~Passkey Authentication in~the~Wild: A Census of~the~Top 100K Sites},
 pages = {319--347},
 publisher = {{Springer Nature Switzerland}},
 isbn = {978-3-032-18268-5},
 editor = {{Ferlin-Reiter, Simone and Fontugne, Romain and Ullrich, Johanna}},
 booktitle = {Passive and Active Measurement},
 year = {2026},
 address = {Cham}
}

@INPROCEEDINGS{Lu.2008,
  author={Lu, H. Karen and Ali, Asad},
  booktitle={Third International Conference on Systems (icons 2008)}, 
  title={Communication Security between a Computer and a Hardware Token}, 
  year={2008},
  volume={},
  number={},
  pages={220-225},
  doi={10.1109/ICONS.2008.36}}

@INPROCEEDINGS{Casagrande.2025,
  author={Casagrande, Marco and Antonioli, Daniele},
  booktitle={2025 IEEE 10th European Symposium on Security and Privacy (EuroS\&P)}, 
  title={CTRAPS: CTAP Client Impersonation and API Confusion on FIDO2}, 
  year={2025},
  volume={},
  number={},
  pages={1034-1048},
  doi={10.1109/EuroSP63326.2025.00063}}

@inproceedings{Sun.2015,
author = {Sun, He and Sun, Kun and Wang, Yuewu and Jing, Jiwu},
title = {TrustOTP: Transforming Smartphones into Secure One-Time Password Tokens},
year = {2015},
isbn = {9781450338325},
publisher = {Association for Computing Machinery},
address = {New York, NY, USA},
url = {https://doi.org/10.1145/2810103.2813692},
doi = {10.1145/2810103.2813692},
booktitle = {Proceedings of the 22nd ACM SIGSAC Conference on Computer and Communications Security},
pages = {976–988},
numpages = {13},
location = {Denver, Colorado, USA},
series = {CCS '15}
}

@article{Ang.2025,
author = {Ang, Kok Wee and Chekole, Eyasu Getahun and Zhou, Jianying},
title = {Unveiling the Covert Vulnerabilities in Multi-Factor Authentication Protocols: A Systematic Review and Security Analysis},
year = {2025},
issue_date = {November 2025},
publisher = {Association for Computing Machinery},
address = {New York, NY, USA},
volume = {57},
number = {11},
issn = {0360-0300},
url = {https://doi.org/10.1145/3734864},
doi = {10.1145/3734864},
journal = {ACM Comput. Surv.},
month = jun,
articleno = {293},
numpages = {37}
}

@book{Rankl.2007,
 author = {Rankl, W. and Cox, K.},
 year = {2007},
 title = {Smart Card Applications: Design models for using and programming smart cards},
 publisher = {Wiley},
 isbn = {9780470511947}
}

@book{Botwright.2023,
 author = {Botwright, R.},
 year = {2023},
 title = {Zero Trust Security: Building Cyber Resilience {\&} Robust Security Postures},
 publisher = {{Pastor Publishing Limited}},
 isbn = {9781839385278}
}

@book{Oorschot.2020,
 author = {{van Oorschot}, P. C.},
 year = {2020},
 title = {Computer Security and the Internet: Tools and Jewels},
 publisher = {{Springer International Publishing}},
 isbn = {9783030336493},
 series = {Information Security and Cryptography}
}

@INPROCEEDINGS{Hsieh.2011,
  author={Hsieh, Wen-Bin and Leu, Jenq-Shiou},
  booktitle={2011 7th International Wireless Communications and Mobile Computing Conference}, 
  title={Design of a time and location based One-Time Password authentication scheme}, 
  year={2011},
  volume={},
  number={},
  pages={201-206},
  doi={10.1109/IWCMC.2011.5982418}}

@article{Aravindhan.2013,
  title={One time password: A survey},
  author={Aravindhan, K and Karthiga, RR},
  journal={International Journal of Emerging Trends in Engineering and Development},
  volume={1},
  number={3},
  pages={613--623},
  year={2013}
}

@inproceedings{Bonneau.2012,
	journal = {2012 IEEE Symposium on Security and Privacy},
	doi = {10.1109/SP.2012.44},
	number = {},
	title = {The Quest to Replace Passwords: A Framework for Comparative Evaluation of Web Authentication Schemes},
	volume = {},
	author = {Bonneau, Joseph and Herley, Cormac and Oorschot, Paul C. van and Stajano, Frank},
	pages = {553--567},
	date = {2012},
	year = {2012},
}

@article{Kumar.2023,
 author = {{Praveen Kumar}, E. and Priyanka, S.},
 year = {2023},
 title = {A password less authentication protocol for multi-server environment using physical unclonable function},
 pages = {21474--21506},
 volume = {79},
 number = {18},
 issn = {1573-0484},
 journal = {The Journal of Supercomputing},
 doi = {10.1007/s11227-023-05437-3}
}

@article{Pahuja.2024,
 author = {Pahuja, Swimpy and Goel, Navdeep},
 year = {2024},
 title = {Multimodal biometric authentication: A review},
 pages = {525--547},
 volume = {37},
 number = {4},
 journal = {AI Communications},
 doi = {10.3233/AIC-220247}
}

\end{document}